\documentclass[useAMS]{gGAF2e}
\usepackage{amsmath}
\usepackage{amssymb}
\usepackage{multiequations}

\newcommand*{\be}{\begin{equation}}
\newcommand*{\ee}{\end{equation}}
\newcommand*{\bea}{\begin{eqnarray}}
\newcommand*{\eea}{\end{eqnarray}}
\newcommand*{\bme}{\begin{multiequations}}
\newcommand*{\eme}{\end{multiequations}}
\newcommand*{\se}{\singleequation}

\newcommand\bftheta{{\mbox{\boldmath$\theta$\unboldmath}}}
\newcommand\bfphi{{\mbox{\boldmath$\phi$\unboldmath}}}

\begin{document}
\doi{10.1080/03091920xxxxxxxxx}
 \issn{1029-0419} \issnp{0309-1929} \jvol{00} \jnum{00} \jyear{2009} 

\title{{\textit{Relativistic Rayleigh-Taylor Instability of a Decelerating Shell
and its Implications for Gamma Ray Bursts}}}

\author{Amir Levinson$^{\ast}$\thanks{$^\ast$Corresponding author. Email: Levinson@wise.tau.ac.il
\vspace{6pt}}\\\vspace{6pt} Raymond and Beverly Sackler School of Physics \& Astronomy, Tel Aviv University,
Tel Aviv 69978, Israel\\\vspace{6pt}\received{v3.3 released February 2009} }

\maketitle

\begin{abstract}
Global linear stability analysis of a self-similar solution describing 
the interaction of a relativistic shell with an ambient medium is performed.  The solution is shown to be 
unstable to convective Rayleigh-Taylor modes having angular scales smaller than the causality scale.  Longer wavelength 
modes are stable and decay with time.  For modes of sufficiently large spherical harmonic degree $l$ the dimensionless 
growth rate scales as $\sqrt{l/\Gamma}$, where $\Gamma$ is the Lorentz factor of the shell.
The instability commences at the contact interface separating the shocked ejecta and shocked ambient gas and 
propagates to the shocks.   The reverse shock front responds promptly to the instability and exhibits rapidly growing
distortions at early times.  Propagation to the forward shock is slower, and it is anticipated that the region near the
contact will become fully turbulent before the instability is communicated to the forward shock.
The non-universality of the Blandford-McKee blast wave solution suggests that turbulence generated by the instability 
in the shocked ambient medium may decay slowly with time and may be the origin of magnetic fields over a long portion of 
the blast wave evolution.   It is also speculated that the instability may affect the emission 
from the shocked ejecta in the early post-prompt phase of GRBs. \bigskip  
\begin{keywords} Relativistic hydrodynamics; Shock waves; Instabilities; Gamma-ray bursts
\end{keywords}\bigskip
\end{abstract}

\section{Introduction}
Relativistic shock waves is a common phenomenon in astrophysics.  They form when a
relativistic flow ejected by a compact central engine interacts with the surrounding medium or, in case
of an intermittent source, as a result of steepening of waves produced in the outflow itself.  The broadband 
emission observed in blazars, micro-quasars and gamma-ray bursts (GRBs) is produced behind those shocks and is 
an important diagnostic of the dissipation process.

Despite an impressive progress in our understanding of relativistic shocks some 
outstanding problems remain open.  Of particular interest is the relativistic blast wave
that form in GRB explosions when the relativistic ejecta expelled by the source impact
the circumburst medium.   The afterglow emission observed in most long GRBs is most likely produced 
in the thin layer enclosed between the forward shock and the ejecta, and is an important diagnostic
of the blast wave evolution and the conditions in the shocked layer.  Although the simple blast wave
model has been quite successful in explaining the late afterglow evolution,
recent observational efforts revealed some features that require extension of the simple model:  
(i) Observations of the late afterglow emission 
indicate strong amplification of magnetic fields in the post shock region - by several orders of
magnitudes larger than what can be achieved by compression of the ambient magnetic field. Despite 
recent efforts to investigate potential mechanisms by which magnetic fields can be generated or amplified
in the vicinity of the shock, this issue remains unresolved.  (ii) SWIFT observations during the early afterglow 
phase reveal strong deviation of the lightcurve at early times from that predicted by the simple blast wave model. 
Several {\em ad hoc} explanations have been offered, including prolonged activity of the central engine 
and evolution of microphysical parameters.  However, the feasibility of these scenarios depends on poorly understood physics, and
it remains to be demonstrated that they can be derived from first principles.
(iii)  In the fireball scenario commonly adopted,
the naive expectation has been that the crossing of the reverse shock should produce an observable 
optical flash.  Despite considerable observational efforts, such flashes seem to be very rare. One
plausible explanation is that the ejecta is magnetically dominated \citep{LE93,LB03,GS05,GMA08}.  Poynting flux 
dominated outflows have the advantage that they can naturally account for the ultra-relativistic Lorentz factors
inferred.   On the other hand, they are challenged by the rapid dissipation of magnetic energy that seems to be required.
Furthermore, even if the flow is magnetically dominated 
some accumulation of baryon rich matter at the 'piston's' head is anticipated during the shock 
breakout phase, that may mimic effects of a hydrodynamic ejecta.

In this paper we explore the stability of the double-shock system.  Hydrodynamic instabilities
can give rise to strong distortions of the structure that may generate turbulence, amplify magnetic fields,
and affect the emission processes in the post-prompt phase.  Such effects have been studied in the 
non-relativistic case in connection with young supernovae remnants (SNRs).
In fact, the idea that the Rayleigh-Taylor (R-T) instability should play an important role in 
the deceleration of a non-relativistic ejecta dates back to Gull \citeyearpar{Gu73}, who 
performed 1D simulations of young SNRs that incorporate a simple model of convection.
\cite{CBE92} later performed a global linear stability analysis of a self-similar solution 
describing the interaction of non-relativistic ejecta with an ambient medium and found that it is subject to a 
convective instability.  They analyzed self-similar perturbations and showed that the flow is unstable
for modes having angular scales smaller than some critical value.
The convective growth rate was found to be largest at the contact discontinuity
surface and to increase with increasing $l$ number of the eigenmodes.  They
also performed 2D hydrodynamical simulations that verified the linear results and 
enabled them to study the nonlinear evolution of the instability.  The simulation
exhibits rapid growth of fingers from the contact interface that saturates, in the nonlinear state,
by the Kelvin-Helmholtz instability.  Strong distortions of the contact and the reverse shock was observed with little
effect on the forward shock.  \cite{JN96} performed 2 and 3D 
MHD simulations of the instability to study the evolution of magnetic fields in the convection
zone.  They confirmed the rapid growth of small scale structure reported in \citep{CBE92}, and in addition 
found strong amplification of ambient magnetic fields in the turbulent flow around R-T 
fingers. On average, the magnetic field energy
density reaches about 0.5\% of the energy density of the turbulence, but it could well be 
that the magnetic field amplification was limited by numerical resolution in their simulations.
The simulations of \cite{CBE92}  and \cite{JN96}  support earlier ideas, that
the clumpy shell structure observed in young (pre-Sedov stage) SNRs such as Tycho, Kepler and Cas A 
is due to the R-T and K-H instability.  

In this paper we extend the linear stability analysis of \cite{CBE92} into the relativistic regime.  A preliminary
account of the model and results is presented in \citep{L09}.
We find that denser ejecta sweeping a lighter ambient gas are subject to the R-T instability also in the relativistic case.
The reason is that in the rest frame of the decelerating contact there is an effective gravitational force which 
is directed outwards, and so in this frame the denser ejecta is 'on top' of the lighter ambient gas.
The stability of a double-shock system has been investigated by \cite{Xi02} using the thin shell approximation.  However,
this study is limited to large scale modes and neglects pressure gradients and, therefore, 
precludes the convective instability.  Thompson \citeyearpar{Th06} pointed out that a magnetized photon-rich 
shell that propagates through a 
dense Wolf-Rayet wind may be subject to the R-T instability.   Using heuristic arguments he examined the conditions under which
the instability develops and estimated the growth rate.  The scaling of the growth rate found below is consistent
with his result.  Gruzinov \citeyearpar{G00} performed a linear stability analysis of 
a Blandford-McKee (BMK) blast wave solution  \citep{BMK76}, and found that the BMK solution is stable but non-universal,
in the sense that some modes decay very slowly as the system evolves.  Furthermore, the onset of oscillations
of an eigenmode of order $l$ has been seen in the simulation once the Lorentz factor evolved to 
$\Gamma <l$.  The conclusion drawn based on Gruzinov's findings 
is that distortion of the shock front at early times may 
cause significant oscillations during a large portion of its evolution.  If the amplitude 
of these oscillations is sufficiently large, and if the same behavior holds in 
the nonlinear regime then this can lead to generation of vorticity in the post shock region 
\citep{GM07,MN07}, and the consequent amplification of magnetic 
fields, as demonstrated recently by Zhang {\it et al.} \citeyearpar{ZM09}.

The plan of the paper is as follows: In section \ref{sec:BE} we derive the basic equations in a general 
form. In  section \ref{sec:unpert} a class of self-similar solutions for the double-shock structure, obtained 
originally by Nakamura and Shigeyama \citeyearpar{NS06}, 
is reviewed.  These are employed as the unperturbed solutions for our analysis. The linear perturbation analysis 
of these solutions is presented in section \ref{sec:pert}. The implications for gamma-ray bursts are discussed in section 
\ref{sec:imp}. We conclude in section \ref{sec:conc}.  Detailed derivation of main results is given in the 
appendices. 

\section{\label{sec:BE}Basic equations}
Consider an unmagnetized fluid, and let $\rho$, $p$, $\tilde{h}$ and $u^\mu$ denote its proper density, pressure, dimensionless 
specific enthalpy, and 4-velocity, respectively. The stress-energy tensor
then takes the form 
\begin{equation}
T^{\mu\nu}=\rho \tilde{h}u^\mu u^\nu-g^{\mu\nu}p,
\end{equation}
where $g^{\mu\nu}$ is the metric tensor.  Neglecting radiative losses, the dynamics of the flow is governed 
by mass and energy-momentum conservation: 
\bme
\label{cont}
\be
\partial_\mu(\rho u^{\mu})=0,
\qquad\qquad\qquad\qquad
\partial_\mu T^{\mu\nu}=0.
\ee
\eme

Using (\ref{cont}a) the different components of (\ref{cont}b) reduce to

\bme
\se
\label{momentum}
\begin{eqnarray}
W\frac{{\rm d}\ln\gamma}{{\rm d}t}+\gamma^{2}\frac{{\rm d}p}{{\rm d}t}&=&\frac{\partial p}{\partial t},\\
\frac{{\rm d}}{{\rm d}t}\ln\left(p/\rho^{\hat{\gamma}}\right)&=&0,\\
\rho\gamma\frac{{\rm d}}{{\rm d}t}(\tilde{h}\gamma{\bf v}_T)+\nabla_T p&=&0.
\end{eqnarray}
\eme
Here $\gamma=u^0$ is the Lorentz factor of the fluid,
${\bf v}_T$ is the tangential component of the 3-velocity, which we express as ${\bf v}=v_r{\widehat {\bf r}}+{\bf v}_T$,
$\hat{\gamma}$ is the adiabatic index, $W=\rho \tilde{h}\gamma^2$, ${\rm d}/{\rm d}t=(u^\mu/u^0)\partial_\mu$ is the convective derivative, and 
\begin{equation}
\nabla_T\equiv {\widehat \bftheta}\frac{1}{r}\frac{\partial}{\partial\theta}+{\widehat\bfphi}\frac{1}{r\sin\theta}
\frac{\partial}{\partial\phi}.
\end{equation}

If the flow passes through a discontinuous shock front, then the solutions of the flow equations
in the upstream and downstream regions are to be matched at the shock surface, which is 
defined by the equation $\psi(x^\mu)\equiv r-R(t,\theta,\phi)=0$. 
Integration of (\ref{cont}a,b) across the surface lead to the jump conditions
\bme
\label{bcrho}
\be
[\rho u^{\mu}]n_\mu=0,
\qquad\qquad\qquad\qquad
\left[T^{\mu\nu}\right]n_\nu=0,
\ee
\eme
where the square brackets denote the difference of the enclosed quantity across the shock front, and
\begin{equation}
n_\mu=\frac{\partial_\mu\psi}{\sqrt{\partial_\mu\psi\partial^\mu\psi}}\label{normal}
\end{equation}
is a 4-vector normal to the shock front.
\begin{figure}
\centering
\includegraphics[width=14cm]{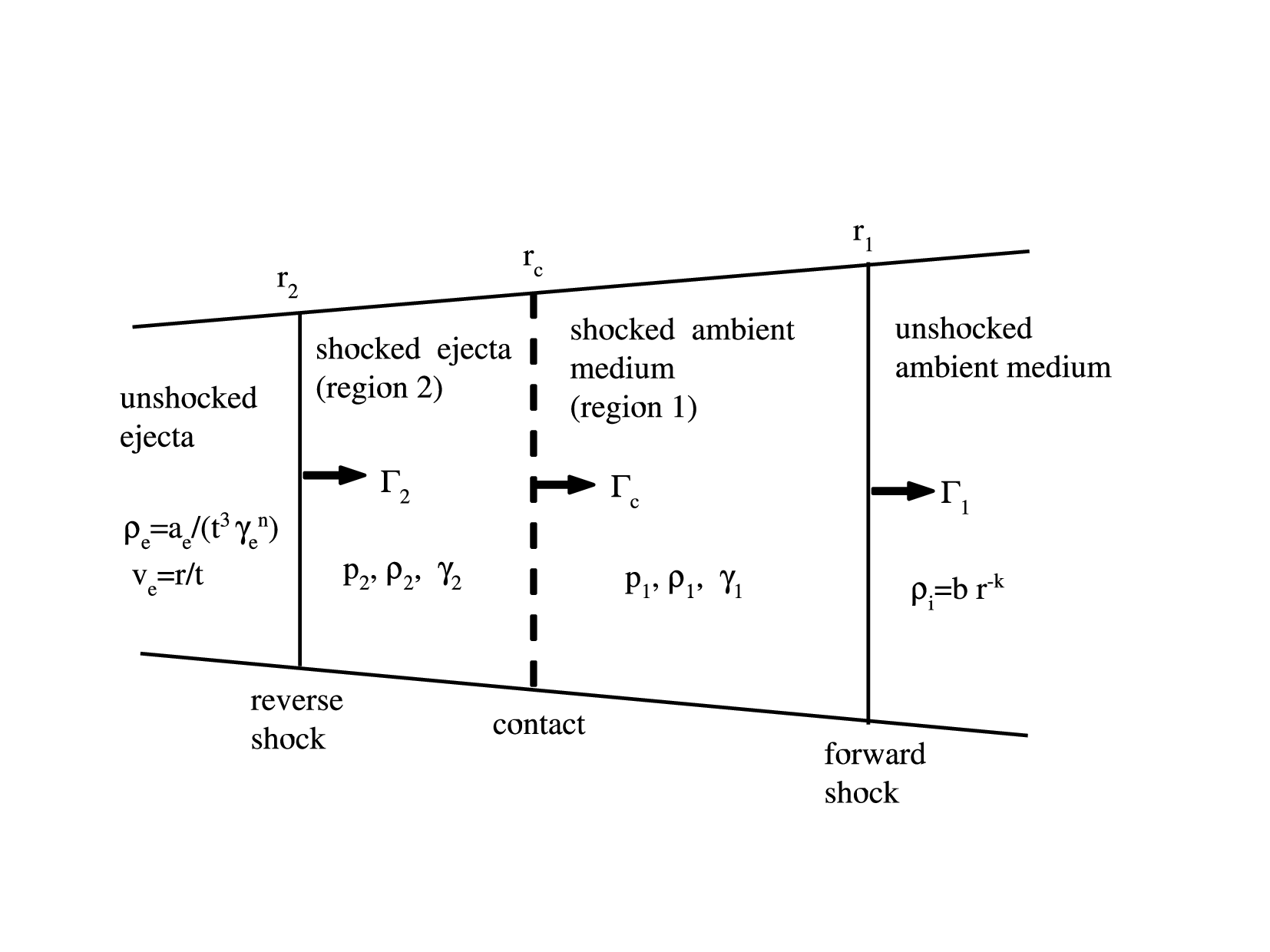}
\caption{\label{fig:a}Schematic representation of the double-shock system.  There are three characteristic surfaces:
a forward shock propagating in the ambient medium, a reverse 
shock sweeping the ejecta, and a contact discontinuity separating the 
shocked ejecta and the shocked ambient medium.  The Lorentz factors of the three surfaces,
measured with respect to the unshocked ambient medium, are indicated.  
Quantities in the shocked ambient medium (region 1) and shocked ejecta (region 2)
are denoted by subscripts 1 and 2, respectively.}
\end{figure}
\section{\label{sec:unpert}Unperturbed solutions}

The unperturbed solution invoked here is the self-similar solution derived by 
Nakamura and Shigeyama \citeyearpar{NS06}.  It is reviewed here to set up the notation 
and to introduce some aspects that are important for the stability analysis. 
Nakamura and Shigeyama considered a freely expanding ejecta interacting
with an ambient medium having a density profile $\rho_i=br^{-k}$.  The freely expanding
ejecta is characterized by a velocity $v_e=r/t$ at time $t$ after the explosion, and  
a proper density profile
\begin{equation}
\rho_e=\frac{a_e}{t^3\gamma_e^n}, \label{den-ejecta}
\end{equation}
where $\gamma_e=1/\sqrt{1-v_e^2}$ is the corresponding Lorentz factor (it can be 
readily seen that the continuity equation (\ref{cont}a) is satisfied for this 
choice of $\rho_e$ and $v_e$). 

The system under consideration is shown schematically in figure \ref{fig:a}.  The subscript 1 refers to 
the shocked ambient medium and 2 to the shocked ejecta.  The Lorentz factors of the
forward shock, reverse shock and the contact discontinuity are denoted by $\Gamma_{1}(t)$,
$\Gamma_{2}(t)$ and $\Gamma_c(t)$, respectively.  Self-similarity requires that they all have a similar 
time evolution, viz., $\Gamma^2_2=At^{-m}$, $\Gamma^2_1=Bt^{-m}$, $\Gamma^2_c=Ct^{-m}$, where 
$A,B,C$ and $m$ are constants determined upon matching 
the solutions in regions 1 and 2 at the contact discontinuity.  The similarity parameter can be defined as \citep{BMK76}
\begin{equation}
\chi=[1+2(m+1)\Gamma_{1}^2](1-r/t)\label{chi}.
\end{equation}
The shocks and the contact are surfaces of constant $\chi$, and since the velocity of a constant 
$\chi$ surface is given by
\begin{equation}
\frac{{\rm d}r}{{\rm d}t}=1-\frac{\chi}{2\Gamma_1^2},
\end{equation}
we readily obtain $\chi_1=1$, $\chi_c=(\Gamma_1/\Gamma_c)^2=B/C>1$ and 
$\chi_2=(\Gamma_1/\Gamma_2)^2=B/A>\chi_c$.  The trajectory of the reverse shock is
\begin{equation}
r_2(t)=\int_0^t{\left(1-\frac{1}{2\Gamma_2^2}\right){\rm d}t^\prime}=t-\frac{t}{2(m+1)\Gamma_2^2},\label{r2}
\end{equation}
from which we obtain for the velocity of the ejecta crossing the shock: $v_e(r_2)=r_2/t=1-1/[2(m+1)\Gamma_2^2]$.
The corresponding Lorentz factor is thus given, to order O($\Gamma_2^{-2}$), by
\bme
\se
\label{g-ejecta}
\be
\gamma_e^2=(m+1)\Gamma_2^2,
\ee
and the density by
\be
\rho_e=\frac{a_e}{t^3\gamma_e^n}=\frac{a_e}{A^{3/m}(m+1)^{n/2}}\Gamma_2^{(6/m)-n}.
\ee
\eme

\subsection{\label{sec:un_1}Shocked ambient medium}
We consider cases where the forward shock is ultra-relativistic.  The specific enthalpy 
of the shocked ambient gas (region 1) is then approximated by $\tilde{h}_1=4p_1/\rho_1$. 
The jump conditions at the forward shock follow from (\ref{bcrho}a,b) and (\ref{normal}) using 
$R(t,\theta,\phi)=r_1(t)=t[1-1/2(m+1)\Gamma_1^2]$.
The self-similar variables for the Lorentz factor, pressure and density are defined as
\bme
\se
\label{g}
\begin{eqnarray}
\gamma_1^2&=&\frac{1}{2}\Gamma_1^2 g(\chi),\\
\rho^\prime_1=\rho_1\gamma_1&=&2\rho_i\Gamma_1^2 h(\chi),\\
p_1&=&\frac{2}{3}\rho_i\Gamma_1^2 f(\chi).
\end{eqnarray}
\eme
Under this choice the shock jump conditions imply $g(1)=f(1)=h(1)=1$.  The equations 
obeyed by the variables $g$, $f$, $h$ are outlined in Appendix \ref{sec:app_unp1}.  Using
 (\ref{g}a) the Lorentz factor of the contact surface can be written as 
$\Gamma_c=\Gamma_1\sqrt{g_c/2}$, where $g_c\equiv g(\chi_c)$, which combined with 
the relation $\chi_c=(\Gamma_1/\Gamma_c)^2$ derived above yields $\chi_c g_c=2$.
The solution is obtained by numerically integrating equations (\ref{out}a-c) 
from the forward shock front $\chi=1$ to the contact discontinuity where $g_c\chi_c=2$.  
The quantities $g_c$, $\chi_c$ are eigenvalues of the solution.

\begin{figure*}
\centering
\includegraphics[width=14cm]{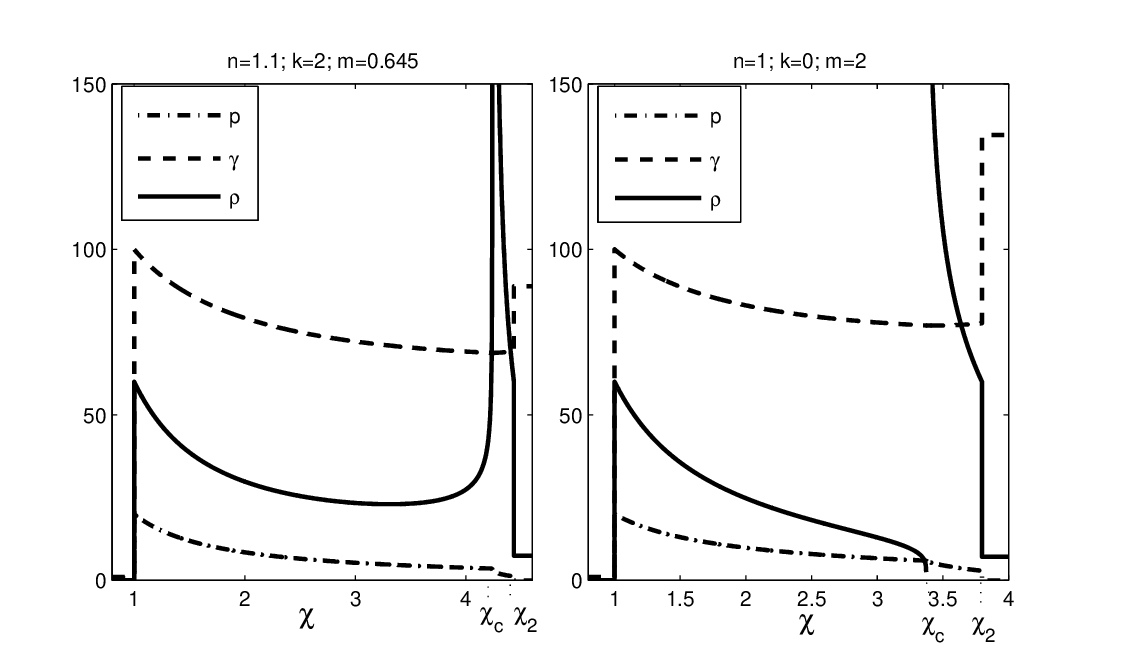}
\caption{\label{fig:b}Profiles of the pressure, proper density and Lorentz factor of the unperturbed flow,
for two different choices of parameters.  The forward
shock is located at $\chi=1$.  The location of the reverse shock ($\chi_2$) and the contact discontinuity ($\chi_c$)
are indicated.}
\end{figure*}
\subsection{\label{sec:un_2}Shocked ejecta}
The reverse shock cannot be considered ultra-relativistic in general and, therefore, a 
complete treatment is required.  The specific enthalpy of the shocked ejecta is 
taken to be $\tilde{h}_2=1+\hat{\gamma}p_2/[\rho_2(\hat{\gamma}-1)]$, and we remind that $\hat{\gamma}$ denotes
the adiabatic index.  Assuming the unshocked ejecta to be cold, the jump conditions at the reverse shock,
obtained from (\ref{bcrho}a,b) and (\ref{normal}) using $R(t,\theta,\phi)=r_2(t)$, 
read
\bme
\se
\label{jump-rho}
\begin{eqnarray}
\rho_e\gamma_e(v_e-V_2)&=&\rho_2\gamma_2(v_2-V_2),\\
\rho_e\gamma^2_e(v_e-V_2)&=&W_2(v_2-V_2)+P_2V_2,\\
\rho_e\gamma^2_ev_e(v_e-V_2)&=&W_2v_2(v_2-V_2)+P_2,
\end{eqnarray}
\eme
here $V_2={\rm d}r_2/{\rm d}t$ is the shock 3-velocity.
Equations (\ref{jump-rho}a-c) can be solved by employing (\ref{g-ejecta}a)
and recalling that $W_2=\rho_2\tilde{h}_2\gamma_2^2$.  One finds
\bme
\se
\begin{eqnarray}
\gamma^2_{2}(\chi_2)&=&q\Gamma_2^2, \\
\rho^\prime_{2}(\chi_2)&=&\frac{mq \rho_e\gamma_e}{(m+1)(q-1)},\\
p_{2}(\chi_2)&=&\frac{m \rho_e}{a(q-1)+2}\left(1-\sqrt{q/(m+1)}\right),
\end{eqnarray}
\eme
where $\sqrt{q}$ is the only positive solution of the equation
\begin{equation}
\hat{\gamma}x^3+(2-\hat{\gamma})\sqrt{m+1}x^2-(2-\hat{\gamma})x-\hat{\gamma}\sqrt{m+1}x=0.\label{q}
\end{equation}

Following Nakamura and Shigeyama \citeyearpar{NS06} we find it convenient to transform in region 2
to a new similarity parameter,
\begin{equation}
\sigma=\chi/\chi_2=\{1+2(m+1)\Gamma_{2}^2\}(1-r/t).\label{sigma}
\end{equation}
The reverse shock is then located at $\sigma=1$ and the contact at 
\begin{equation}
\sigma_c=\chi_c/\chi_2=\Gamma^2_2/\Gamma_c^2. \label{sigc}
\end{equation}
The self similar variables of the shocked ejecta,
$G$, $F$, $H$, are then defined as
\bme
\se
\label{G2}
\begin{eqnarray}
\gamma^2_{2}&=&q\Gamma_2^2 G(\sigma),\\ 
\rho^\prime_{2}&=&\frac{mq \rho_e\gamma_e}{(m+1)(q-1)} H(\sigma),\\
p_{2}&=&\frac{m \rho_e}{a(q-1)+2}\left[1-\sqrt{q/(m+1)}\right]F(\sigma),
\end{eqnarray}
\eme
and satisfy $F(1)=G(1)=H(1)=1$.  From equations (\ref{sigc}) and (\ref{G2}a) we obtain 
$G_c\sigma_c=1/q$ at the contact discontinuity.
The equations obeyed by these self-similar variables are derived
in Appendix \ref{sec:app_unp2}.  The solution in this region is obtained upon integration of 
(\ref{in}a-c) from the reverse shock $\sigma=1$ to the contact $G_c\sigma_c=1/q$.

\subsection{\label{sec:un_contact}Conditions at the contact surface}
Two conditions at the contact discontinuity fix the constants $A$, $B$, and $m$. One condition is 
that there be no flow across the contact interface.  This implies $\Gamma_c=\gamma_{1c}
=\gamma_{2c}$, from which we obtain
\begin{equation}
\frac{\Gamma_1^2}{\Gamma_2^2}=\frac{B}{A}=2q\frac{G_c}{g_c},
\end{equation}
where equations (\ref{g}a) and (\ref{G2}a) have been employed.  The second 
condition is pressure balance, viz., $p_1(t,\chi_c)=p_2(t,\sigma_c)$.  This 
condition yields two relations.  The first one,
\begin{equation}
m=\frac{6-2k}{n+2},\label{m}
\end{equation}
comes from the requirement that $p_1$ and $p_2$ have the same time dependence.
The second one is implied by equations (\ref{g}c) and (\ref{G2}c):
\begin{eqnarray}
\frac{a_e}{bA^{1+n/2}}=\frac{4q[\hat{\gamma}(q-1)+2(\hat{\gamma}-1)]}{3m(\hat{\gamma}-1)}\frac{G_c f_c}{g_c F_c}
\left(\frac{m+1}{\sigma_c}\right)^{n/2}\left(1-\sqrt{\frac{q}{m+1}}\right)^{-1}.
\end{eqnarray}

The solution described above is valid for $-1<m<3-k$ \citep{BMK76}.  The case $m=3-k$ corresponds to an adiabatic impulsive 
blast wave (for which $g\chi=1$ so that the contact surface is undefined).  From (\ref{out}a-c) 
it can be shown \citep{BMK76} that near the contact discontinuity the density of the ambient medium 
behaves as $h\propto(2-g\chi)^{-\theta_1}$, with $\theta_1=(m-k)/(m+3k-12)$. 
Thus, within the range of parameters for which the solution is valid $h$ diverges at the contact for $m<k$ and vanishes for $m>k$.  
Likewise, from  (\ref{in}a-c) we find that the density of the shocked ejecta behaves 
as $H\propto(qG\sigma-1)^{-\theta_2}$, where 
$\theta_2=(1-\hat{\gamma})(mn-6)/(10\hat{\gamma}+2mn-12)=(\hat{\gamma}-1)(6+nk)/[5\hat{\gamma}(n+2)-2nk-12]$.  Within the allowed
range of parameters $\theta_2$ is always positive, so that $H$ always diverges.  This is seen in 
figure \ref{fig:b}, where solutions obtained for an ejecta interacting with a uniform density medium (right panel) 
and a stellar wind (left panel) are exhibited.

\section{\label{sec:pert}Global linear stability analysis}
We now consider linear perturbations of the self-similar solution described above. As shown below,
unlike in the non-relativistic case \citep{CBE92} the relativistic perturbation equations do not admit
self-similar solutions, with the exception of the spherical mode.  The reason is the inherent coupling, via the Lorentz factor,
of the radial and tangential velocity perturbations $\delta v_r$ and $\delta v_T$.  Thus, numerical integration of
the time dependent equations is required. 
\subsection{\label{sec:pert_eq}Perturbation equations}
The perturbed variables are taken to be 
\bme
\se
\label{delrho}
\begin{eqnarray}
\rho^\prime&=&\rho^\prime_0(t,r)+\delta\rho^\prime(t,r,\theta,\phi),\\
p&=&p_0(t,r)+\delta p(t,r,\theta,\phi),\\
{\bf v}&=&v_0(r)\hat{\bf r}+\delta v_r(r,\theta,\phi)\hat{\bf r}+\delta{\bf v}_T(r,\theta,\phi),
\end{eqnarray}
\eme
here the subscript zero denotes the unperturbed flow.
To simplify the notation we define ${\rm d}^0_t=\partial_t+v_0\partial_r$ to be the 
convective derivative with respect to the unperturbed flow. The Lagrangian change
of some quantity $Q$ is then given to first order by ${\rm d}Q/{\rm d}t={\rm d}_t^0(Q_0+\delta Q)+\delta v_r\partial _r Q_0$.
The linearized equations for the perturbations, obtained upon substitution of the expressions (\ref{delrho}a-c) 
into the hydrodynamic equations (\ref{momentum}a-c) are
\bme
\se
\label{prt}
\begin{eqnarray}
{\rm d}^0_t(\delta\rho^\prime/\rho_0^\prime)+\partial_r \delta v_r
+(2/r+\partial_r\ln\rho_0^\prime) \delta v_r+\nabla(\delta{\bf v}_T)&=&0,\\
{\rm d}_t^0\left(\frac{\delta p}{p_0}-\hat{\gamma}\frac{\delta \rho^\prime}{\rho^\prime_0}
+\hat{\gamma}\frac{\delta \gamma}{\gamma_0}
\right)+\delta v_r\partial_r \ln(p_0\rho_0^{-\hat{\gamma}})&=&0,\\
\rho_0^\prime {\rm d}^0_t(\tilde{h}_0\gamma_0\delta{\bf v}_T)+\nabla_T \delta p&=&0,\\
W_0\left[{\rm d}_t^0(\gamma_0^2\delta v_r)+\delta v_r\partial_r\ln\gamma_0\right]
+\delta W {\rm d}_t^0(\ln\gamma_0)&& \nonumber\\ 
+\gamma^2_0{\rm d}_t^0\delta p+\gamma_0^2\delta v_r\left[2\gamma_0^2{\rm d}_t^0p_0+\partial_rp_0\right]
-\partial_t\delta p&=&0.
\end{eqnarray}
\eme
Next, we apply the perturbation equations (\ref{prt}a-d) to regions 1 and 2 (see figure \ref{fig:a}), using
the self-similar solutions in each region for the unperturbed quantities.  

\subsubsection{\label{sec:pert_eq_1}Perturbed flow in region 1}
We expand the perturbations in spherical harmonics and use
$\chi$ and $\tau=\ln t$ as new independent variables in place of $r,t$.  
The perturbations in region 1 are then expressed as
\bme
\se
\label{xirho}
\begin{eqnarray}
\delta \rho^\prime_1&=&\rho_{10}^\prime\xi_\rho(\tau,\chi)Y_{l\tilde{m}}(\theta,\phi),\\
\delta p_1&=&p_{10}\xi_P(\tau,\chi)Y_{l\tilde{m}}(\theta,\phi),\\
\delta v_{1r}&=&\frac{1}{\Gamma_1^2 g}\xi_R(\tau,\chi)Y_{l\tilde{m}}(\theta,\phi),\\ 
\delta{\bf v}_{1T}&=&\xi_T(\tau,\chi)r\nabla_T Y_{l\tilde{m}}(\theta,\phi),
\end{eqnarray}
\eme
with $\rho_{10}^\prime$ and $p_{10}$ given by (\ref{g}b,c), respectively.
When these expressions are substituted in equations (\ref{prt}a-d) a set of first order 
hyperbolic PDEs for the dimensionless amplitudes $\xi_\alpha$ is obtained:
\begin{equation}
\partial_\tau\xi_\alpha=\Sigma_\beta\{A_{\alpha\beta}\partial_\chi\xi_\beta.
+B_{\alpha\beta}\xi_\beta\},\label{xiset}
\end{equation}
where the indices $\alpha,\beta$ run over $R,P,\rho,T$.  The details are given in Appendix \ref{sec:app_r1}.
The coefficients $A_{\alpha\beta}$ and $B_{\alpha\beta}$, given 
explicitly in (\ref{ARR}a-c) and (\ref{BRR}a-g), are functions
of the self-similarity coordinate $\chi$, but are independent of $\tau$.

\subsubsection{\label{sec:pert_eq_2}Perturbed flow in region 2}
Likewise, in region 2 we transform to the coordinates $\sigma, \tau$ and define 
\bme
\se
\label{etarho}
\begin{eqnarray}
\delta \rho^\prime_2&=&\rho_{20}^\prime\eta_\rho(\tau,\sigma)Y_{l\tilde{m}}(\theta,\phi),\\
\delta p_2&=&p_{20}\eta_P(\tau,\sigma)Y_{l\tilde{m}}(\theta,\phi),\\
\delta v_{2r}&=&\frac{1}{\Gamma_2^2 G}\eta_R(\tau,\sigma)Y_{l\tilde{m}}(\theta,\phi),\\ 
\delta{\bf v}_{2T}&=&\eta_T(\tau,\sigma)r\nabla_T Y_{l\tilde{m}}(\theta,\phi).
\end{eqnarray}
\eme
The unperturbed parameters of the shocked ejecta, $\rho_{20}^\prime$, $p_{20}$ are given by
equations (\ref{G2}b,c).  The derivation of the perturbation equations for the dimensionless
perturbations $\eta_\alpha$ is presented in Appendix \ref{sec:app_r2}. The resultant set of equations is

\begin{equation}
\partial_\tau\eta_\alpha=\Sigma_\beta\{C_{\alpha\beta}\partial_\chi\eta_\beta
+D_{\alpha\beta}\eta_\beta\},\label{etaset}
\end{equation}
with the coefficients $C_{\alpha\beta}(\sigma)$ and $D_{\alpha\beta}(\sigma)$ given 
explicitly in (\ref{C1}a-d) and (\ref{DTT}a-h).

\subsection{\label{sec:pert_bc}Boundary conditions}
Equations (\ref{xiset}), (\ref{etaset}) are solved subject to boundary conditions imposed at the shock fronts
and at the contact discontinuity.  We allow perturbations of the shock fronts and the 
contact surface of the form
\begin{equation}
\delta r_a(t,\theta,\phi)=\frac{t\delta_a(t)}{\Gamma_a^2}Y_{l\tilde{m}}(\theta,\phi),\label{del_rs}
\end{equation}
here $a=1,2,c$ refers to the forward shock, reverse shock, and the contact discontinuity, respectively.
The corresponding perturbation of the 3-velocity at these surfaces is
\begin{equation}
\delta V_a=\Gamma_a^{-2}[\partial_\tau\delta_a+(m+1)\delta_a]Y_{l\tilde{m}}(\theta,\phi).\label{delVj}
\end{equation}
Now, the Lagrange perturbation of some fluid quantity $Q$ at the perturbed position of surface $a$ is given by
$\Delta_a Q=(\partial_r Q_0)\delta r_a +\delta Q$; e.g.,  
$\Delta_a \rho^\prime=(\partial_r \rho^\prime_0)\delta r_a +\delta \rho^\prime$, etc.  Defining $v^\mu=u^\mu/u^0$
and denoting by $n_{j\mu}=n^0_{j\mu}+\delta n_{j\mu}$ the perturbed normal of the forward ($j=1$)/reverse ($j=2$) 
shock, equation (\ref{bcrho}a) gives to first order
\bme
\se
\label{cont-s}
\be
[(\rho_0^\prime+\Delta_j \rho^\prime)(v^\mu_0+\Delta_j v^\mu)(n^0_{j\mu}+\delta n_{j\mu})]=0,\\
\ee
and (\ref{bcrho}b)
\be
[(W_0+\Delta_j W)(v^\nu_0+\Delta_j v^\nu)(v^\mu_0+\Delta_j v^\mu)
-(p_0+\Delta_j p)g^{\nu\mu}](n^0_{j\mu}+\delta n_{j\mu})=0.
\ee
\eme
The forward shock is described by the equation 
$\psi_{1}(x^\mu)=r-r_{1}(t)-\delta r_{1}(t,\theta,\phi)$, from which we obtain, using (\ref{normal}),
\bme
\se
\label{pert_normal}
\begin{eqnarray}
n_{1\mu}^{0}&=&(-\Gamma_{1}V_{1},\Gamma_{1},0),\\
\delta n_{1\mu}&=&(-\Gamma^3_{1}\delta V_{1},\Gamma^3_{1} V_{1}\delta V_{1},
-\Gamma_{1}\nabla_T\delta r_{1}).
\end{eqnarray}
\eme
The density, pressure, and Lorentz factor of the unshocked flow just upstream of the forward shock
are $p_i=0$, $\rho_i=br^{-k}$, and $\gamma_i=1$.  At the perturbed shock front 
$\Delta_1\rho_i=-k\rho_i\delta r_1/r_1$ and can be neglected to the order to which we are working.
Applying (\ref{xirho}a-d) to the forward shock and using (\ref{cont-s}a), (\ref{pert_normal}a,b) we get
\bme
\se
\label{bc-xirho}
\be
\xi_\rho=\partial_\tau\delta_1+2\xi_R+(3m-3+2k)\delta_1.
\ee
Likewise, the transverse component of (\ref{cont-s}b) yields
\be
\xi_T=-\delta_1/\Gamma_1^2,
\ee
and the other two components
\begin{eqnarray}
\xi_P&=&2\partial_\tau\delta_1+(14-8m-6k)\frac{\delta_1}{3},\\
\xi_R&=&2\partial_\tau\delta_1-2(m-3+k)\delta_1.
\end{eqnarray}
\eme
Finally, we eliminate $\delta_1$ from equations (\ref{bc-xirho}c,d) to get three boundary conditions
at the forward shock ($\chi=1$):
\bme
\se
\label{bc-xi}
\begin{eqnarray}
\partial_\tau\xi_P-\partial_\tau\xi_R&=&(m-3+k)\xi_P+\frac{1}{3}(7-4m-3k)\xi_R,\\
\xi_T&=&\frac{3}{2(m+2)\Gamma_1^2}[\xi_P-\xi_R],\\
\xi_\rho&=&-\partial_\tau(\Gamma_1^2\xi_T)+2\xi_R-(3m-3+2k)\Gamma_1^2\xi_T.
\end{eqnarray}
\eme
As a check note that for the impulsive BMK solution with $m=3$ and $k=0$ equations (\ref{bc-xi}a,b)
reduce to those derived in \citep{G00}.

The derivation of the boundary conditions at the reverse shock is far more involved. The details are 
presented in Appendix \ref{sec:app_BC}. One finds
\bme
\se
\label{bc-eta}
\begin{eqnarray}
\partial_\tau\eta_R&=&-\frac{d_{sP}-f_{sP}d_{s\rho}}{d_{sR}-f_{sR}d_{s\rho}}\partial_\tau\eta_P
-\frac{1-f_{s\delta}d_{s\rho}}{d_{sR}-f_{sR}d_{s\rho}}\partial_\tau\delta_2,\\
\partial_\tau\eta_\rho&=&-f_{sR}\partial_\tau\eta_R-f_{sP}\partial_\tau\eta_P
-f_{s\delta}\partial_\tau\delta_2,\\
\eta_T&=&-\frac{2}{\kappa(q-1)\Gamma_2^2}[d_{sR}\eta_R+d_{sP}\eta_P+d_{s\rho}\eta_{\rho}],\\
\partial_\tau\delta_2&=&-e_{sR}\eta_R-e_{sP}\eta_P-e_{s\rho}\eta_{\rho},
\end{eqnarray}
\eme
at $\sigma=1$.  The coefficients are given in (\ref{cof-2}a-c) of Appendix \ref{sec:app_BC}.  

Two additional boundary conditions are imposed at the contact discontinuity. The requirement that there be no flow across the 
contact surface, that is $v-{\rm d}r_c/dt=0$, implies $\partial_r v_0\delta r_c+\delta v_r-{\rm d}^0_t\delta r_c=0$ on each side 
of that surface.  Upon substitution of the unperturbed solution we obtain
\bme
\se
\label{bc_p_cont}
\begin{eqnarray}
\partial_\tau\delta_c&=&\frac{\xi_R(\tau,\chi_c)}{2}-(m+1)\left[1+\chi_c(\partial_\chi\ln g)_c\right]
\delta_c,\\
\partial_\tau\delta_c&=&q\eta_R(\tau,\sigma_c)-(m+1)\left[1+\sigma_c
(\partial_\sigma\ln G)_c\right]\delta_c,
\end{eqnarray}
where subscript c refers to values at the contact.  Pressure balance across the contact discontinuity yields
\be
\eta_P(\tau,\sigma_c)-\xi_P(\tau,\chi_c)=2(m+1)[\sigma_c(\partial_\sigma\ln F)_c
-\chi_c(\partial_\chi\ln f)_c]\delta_c.
\ee
\eme
After some manipulation of (\ref{bc_p_cont}a-c) we finally arrive at
\bme
\se
\label{bc_p_cont4}
\begin{eqnarray}
\partial_\tau\xi_R(\tau,\chi_c)-2q\partial_\tau\eta_R(\tau,\sigma_c)&=&
-(m+1)[1+\sigma_c(\partial_\sigma\ln G)_c]\xi_R(\tau,\chi_c)\nonumber\\
&&\,+2q(m+1)[1+\chi_c(\partial_\chi\ln g)_c]\eta_R(\tau,\sigma_c),\\
\eta_P(\tau,\sigma_c)-\xi_{P}(\tau,\chi_c)&=&
2(m+1)\{\sigma_c\partial_\sigma\ln F)_c
-\chi_c\partial_\chi\ln f)_c\}\delta_c,
\end{eqnarray}
with
\begin{equation}
\delta_c(\tau)=\frac{\xi_R(\tau,\chi_c)-2q\eta_R(\tau,\sigma_c)}{2(m+1)[\chi_c(\partial_\chi\ln g)_c
-\sigma_c(\partial_\sigma\ln G)_c]}.
\end{equation}
\eme
Equations (\ref{bc-xi}a-c), (\ref{bc-eta}a-d) and (\ref{bc_p_cont4}a-c) provide 
a set of eight boundary conditions for the perturbation equations.

\subsection{\label{sec:pert_scheme}Numerical scheme}
To integrate equations (\ref{xiset}) and (\ref{etaset}) we first transform to a new set of variables, 
the so called ``Riemann invariants''.  In region 1 those are related to the old variables through
\bme
\label{Rm-out}
\be
\se
\xi_\pm=\frac{1}{\sqrt{3}}\xi_R\pm\frac{1}{2}\xi_P,
\ee
\be
\xi_3=\xi_{\rho}-\xi_R/2-3\xi_P/4,
\qquad\qquad\qquad\qquad
\xi_4=\xi_T,
\ee
\eme
and in region 2 through
\bme
\label{Rm-in}
\be
\se
\eta_\pm=\frac{\eta_R}{2}\pm\frac{1}{2q\sqrt{\kappa\hat{\gamma}}}\eta_P,
\ee
\be
\eta_3=\eta_{\rho}-q\eta_R-\eta_P/\hat{\gamma}, 
\qquad\qquad\qquad\qquad
\eta_4=\eta_T,
\ee
\eme
with $\kappa(\sigma)=W_{20}/p_{20}\gamma_{20}^2$, as
defined in Appendix \ref{sec:app_unp2}.  The equations for the new variables can be obtained upon appropriate
transformations of  (\ref{xiset}), (\ref{etaset}), and are derived in Appendix \ref{sec:app_RI}.  As shown 
there, $\xi_-,\xi_3,\xi_4$ propagate from the forward shock to the contact discontinuity while $\xi_+$ propagates 
in the opposite direction.  Likewise, $\eta_+,\eta_3,\eta_4$ propagate from the reverse shock to the 
contact discontinuity whereas $\eta_-$ propagates from the contact discontinuity to the reverse shock.  Thus, the two boundary conditions
at the contact discontinuity, (\ref{bc_p_cont4}a,b), are applied to $\xi_+$ 
and $\eta_-$ after appropriate transformation of variables, and the boundary conditions at the forward and reverse shocks
to the remaining Riemann invariants.
 
Numerical simulations of equations (\ref{Riemn1}) and (\ref{Riemn2}) were performed using the forward Euler
scheme with upwind differencing of spatial derivatives.  We also used for comparison a 4th order Runge-Kutta routine 
with an adaptive step size to advance the equations in time and found unnoticeable differences between 
the two methods for all cases studied below. 
For each experiment we have made several runs with increasing grid resolution until the result
converged.  The divergence of the unperturbed density at the contact have caused no difficulties. 
To test the code we exploited the analytic solutions for the spherical modes obtained
below.  We generally found excellent agreement.  As an example, for the choice $n=1.1, k=2$
for which $m=0.645$, and with $10^4$ grid points on each side of the contact discontinuity all of the
Riemann invariants followed the analytic solution to an accuracy better than $10^{-4}$ up to a time 
$\tau=5$ (or $t/t_0={\rm e}^5$).  The accuracy of the initial condition, specifically the 
accuracy at which the boundary conditions at the contact at $\tau=0$ were matched was $4.8\times10^{-5}$ 
in this run.  

As a second test we solved (\ref{Riemn1}) for an impulsive BMK solution with $k=0$.  For this solution a contact 
surface does not exist ($g\chi=1$ for every $\chi$), so the boundary condition for $\xi_+$ at the contact 
needs to be replaced.  We verified that the solution depends weakly on this condition provided it is fixed
sufficiently far from the forward shock, at $\chi\gg 1$.   We compared our result with that obtained in \cite{G00} and
found excellent agreement\footnote[1]{There is a small error in Gruzinov \citeyearpar{G00}. The coefficient of the 3rd 
term on the right-hand side
of his  30 should be -17/3 (see  (\ref{bmk}c)) and not -14/3 as in the astro-ph/0012364 version.  However, 
we have confirmed that the effect of this error on the result is insignificant.}.

\subsection{\label{sec:pert_sph}Spherical perturbations}
For spherical perturbations ($l=0$) the tangential velocity vanishes, viz., $\delta {\bf v}_T=0$, as can be 
directly seen from  (\ref{xirho}d) and (\ref{etarho}d).  Then (\ref{xiset}), (\ref{etaset})
admit solutions of the form $\xi_{\alpha}(\chi,\tau)=\xi_\alpha(\chi){\rm e}^{s\tau}$,
$\eta_{\alpha}(\sigma,\tau)=\eta_\alpha(\sigma){\rm e}^{s\tau}$; $\alpha=(R,P,\rho)$, 
where $s$ is an eigenvalue determined from the boundary conditions at the contact discontinuity.
There are two spherical modes.  One is associated with a linear time translation of the unperturbed
solution; that is, it is the difference between a solution at time $t$ and a solution at time $t+\delta t$.
Since to lowest order all surfaces propagate at the speed of light the distance between corresponding surfaces
of the two solutions should remain $\delta t$ at all times.  This implies $\delta V_a=0$ to order ${\rm O}(\Gamma ^{-2})$ 
and from (\ref{delVj}) we anticipate
that $s=-(m+1)$ for this mode.  From (\ref{chi}) and (\ref{sigma}) it can be readily seen that the change 
in the self-similarity parameters $\chi$ and $\sigma$ corresponding to a time translation $\delta t$ 
is $\delta\chi=2(m+1)\Gamma_1^2\delta t/t=2(m+1)B{\rm e}^{-(m+1)\tau}\delta t$ and $\delta \sigma =\delta \chi/\chi_2$.
To order ${\rm O}(\Gamma_1^{-2})$ equations (\ref{g}a-c)) yield $\delta \gamma_1^2
=(\partial_\chi\gamma_1^2)\delta \chi$, $\delta p_1=(\partial_\chi p_{10})\delta \chi$, and 
$\delta \rho^\prime_1=(\partial_\chi \rho^\prime_{10})\delta \chi$.
Note that since $\delta\chi\partial_\chi=-\delta t\partial_r$ (see (\ref{dt2}b)) the latter relations
simply mean that the Lagrange perturbations of the flow parameters vanish; that is,
$\Delta p_1=\delta p_1+\delta t\partial_r p_{10}=\delta p_1-(\partial_\chi p_{10})\delta \chi=0$, etc.
By employing  (\ref{xirho}a-c) we finally obtain
\bme
\se
\label{spher-mod-1}
\begin{eqnarray}
\xi_R(\chi)&=&2(m+1)\partial_\chi \ln g,\\
\xi_p(\chi)&=&2(m+1)\partial_\chi \ln f,\\
\xi_\rho(\chi)&=&2(m+1)\partial_\chi \ln h,
\end{eqnarray}
\eme
in region 1.  Likewise, in region 2 we find, using (\ref{G2}a-c)
\bme
\se
\label{spher-mod-2}
\begin{eqnarray}
\eta_R(\chi)&=&\frac{(m+1)}{q}\partial_\chi \ln G,\\
\eta_p(\chi)&=&2(m+1)\partial_\chi \ln F,\\ 
\eta_\rho(\chi)&=&2(m+1)\partial_\chi \ln H.
\end{eqnarray}
\eme
By direct substitution it can be shown that $\xi_\alpha(\tau,\chi)=\xi_\alpha(\chi){\rm e}^{-(m+1)\tau}$,
$\eta_\alpha(\tau,\chi)=\eta_\alpha(\chi){\rm e}^{-(m+1)\tau}$, with $\xi_\alpha(\chi)$ and $\eta_\alpha(\chi)$
given by (\ref{spher-mod-1}) and (\ref{spher-mod-2}), respectively,  
satisfy the perturbation equations (\ref{xiset}), (\ref{etaset}), and all the boundary conditions. Note
that for this mode $\delta r_{20}=\delta r_2$ in (\ref{Asdel}b) and (\ref{Bsdel}e). 

The second eigenmode of order $l=0$ was found by numerically integrating the set of ODEs obtained upon substitution of the 
relations $\xi_{\alpha}(\chi,\tau)=\xi_\alpha(\chi){\rm e}^{s\tau}$,
$\eta_{\alpha}(\sigma,\tau)=\eta_\alpha(\sigma){\rm e}^{s\tau}$ into equations (\ref{xiset}) and (\ref{etaset}).
The method of solution was to guess the eigenvalue $s$ and integrate the equations in each region from the shock
to the contact surface.  The process was repeated until the condition (\ref{bc_p_cont4}b) was satisfied to the required
accuracy.  For each choice of parameters we found, using this method, two solutions; the first one coincides with the analytic solution
given by (\ref{spher-mod-1}) and (\ref{spher-mod-2}).  The second one has a smaller decay rate, $s>-(m+1)$.  

\begin{figure*}
\centering
\includegraphics[width=14cm]{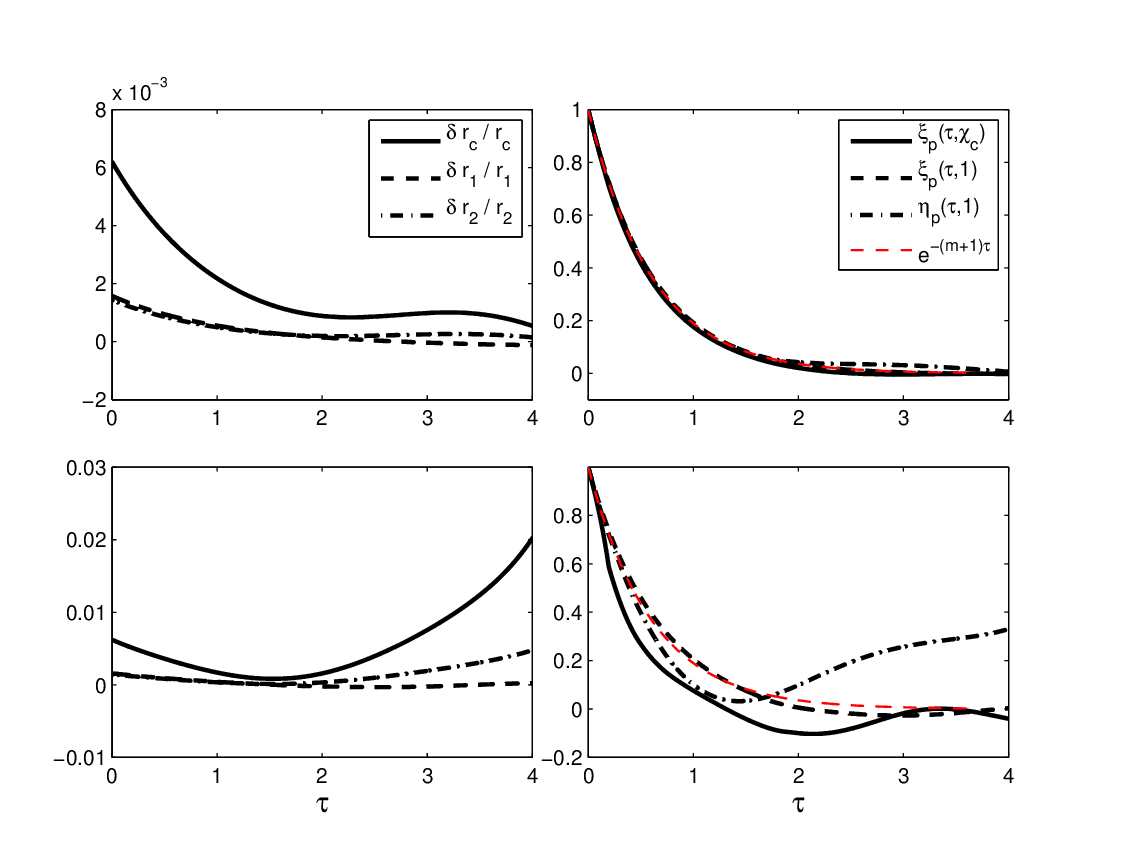}
\caption{\label{fig:c}Time evolution of the perturbations for $n=1.1$, $k=2$, $l(l+1)/\Gamma_{10}^2=0.2$
(upper panels) and $l(l+1)/\Gamma_{10}^2=1.5$ (lower panels). Here $\Gamma_{10}$ being the initial Lorentz factor of 
the forward shock.  The dimensionless distortions of the different surfaces are delineated in the left
panels and the relative pressure perturbations (i.e., normalized to their initial values) in the right panels, as indicated.
The dashed red line corresponds to the analytic solution of the spherical mode and is plotted here for a comparison.}
\end{figure*}
\begin{figure*}
\centering
\includegraphics[width=16cm]{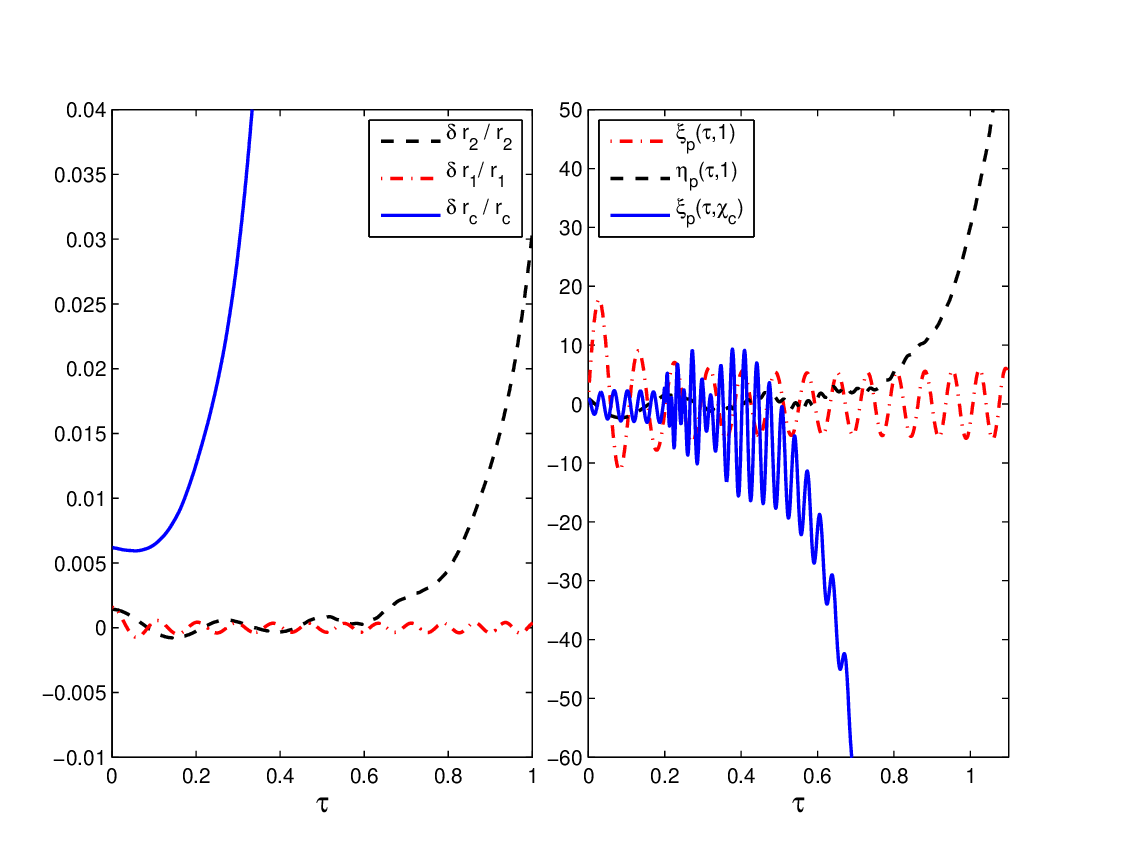}
\caption{\label{fig:d}Same as figure 2 but for $l(l+1)/\Gamma_{10}^2=10^4$. Left panel: dimensionless distortions
of the forward (dotted-dashed line) and reverse (dashed line) shock fronts and the contact surface (solid line).  Right panel:
relative pressure perturbations at the forward shock front (dotted-dashed line), reverse shock front (dashed line), and 
the contact surface (solid line).}
\end{figure*}
\begin{figure*}
\centering
\includegraphics[width=14cm]{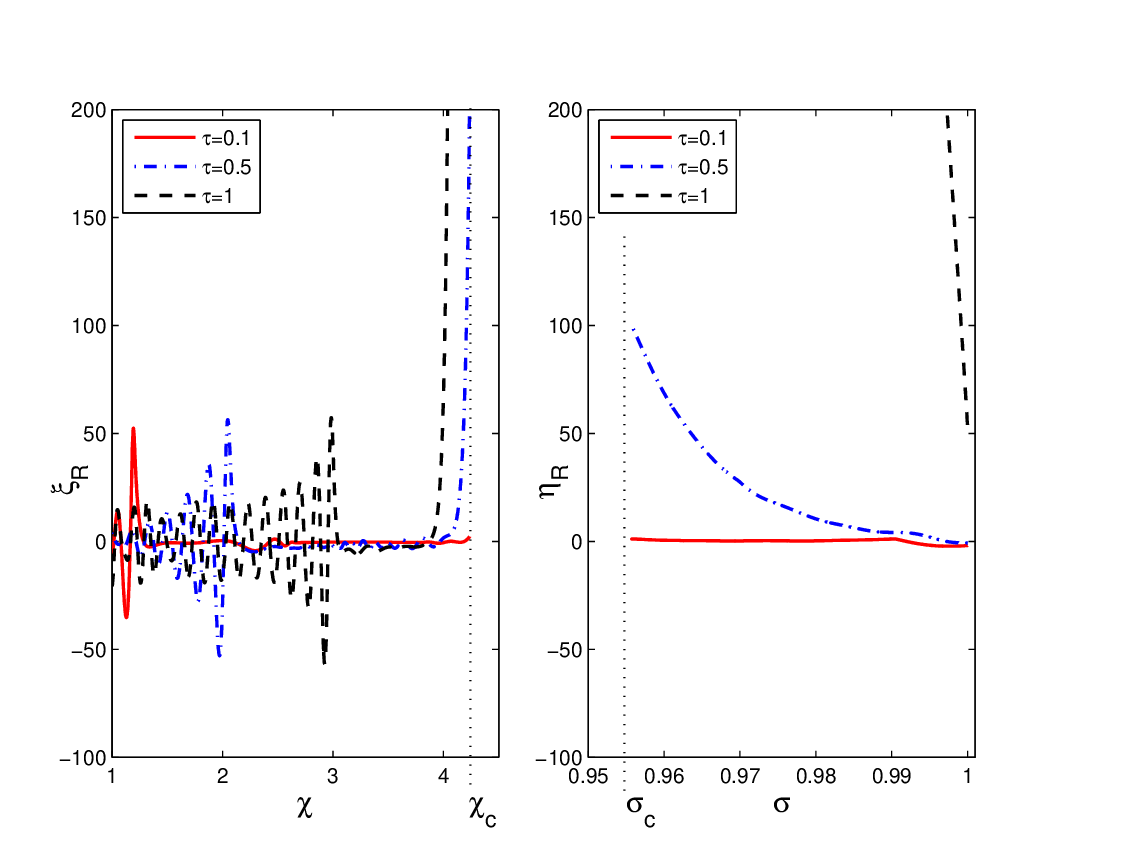}
\caption{\label{fig:e} Profile of the dimensionless perturbation of radial velocity at different times.  
Left panel: shocked ambient medium (region 1). 
The location of the forward shock is at $\chi=1$.  The location of the contact discontinuity $\chi_c$ is indicated.
Rapid growth of the perturbation near the contact is clearly seen.  The wave propagating from the forward shock to the 
contact interface corresponds to transmission of the initial shock disturbance.
Right panel:  shocked ejecta (region 2).   Here $\eta_R$ is plotted against the normalized variable $\sigma=\chi/\chi_2$
(see equation (\ref{sigma})).  The locations of the reverse shock and the contact discontinuity are at $\sigma=1$ 
and $\sigma=\sigma_c=\chi_c/\chi_2$, respectively.  As seen, at $\tau=1$ the instability already propagated throughout the 
entire region.} 
\end{figure*}
\subsection{\label{sec:pert_nonsph}Non-spherical perturbations}
\begin{figure}
\centering
\includegraphics[width=9cm]{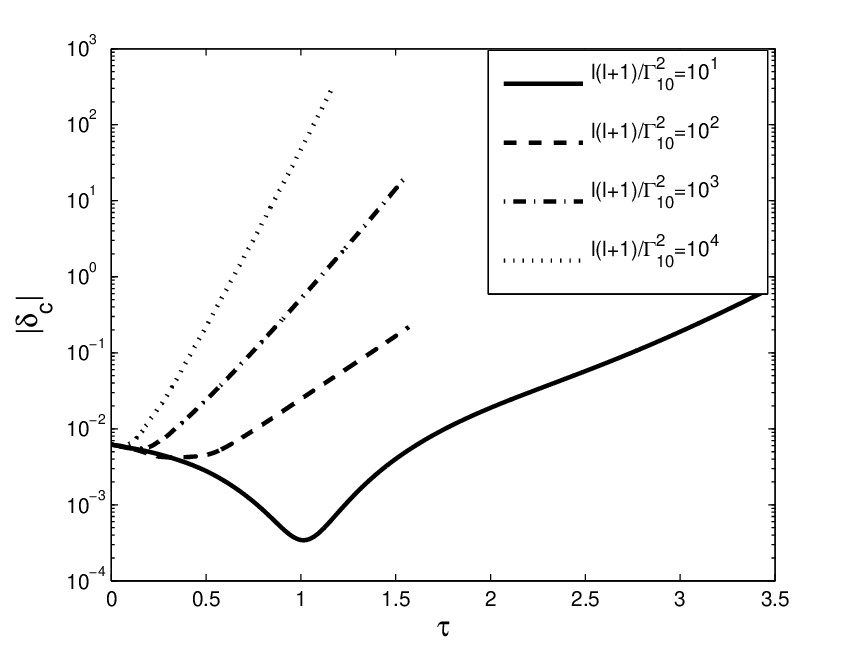}
\caption{\label{fig:f}Absolute value of the relativistic distortion $\delta_c$ defined in equation (\ref{del_rs}) (see also
(\ref{bc_p_cont4}c)) versus time $\tau$, for different values of the spherical harmonic degree $l$.}
\end{figure}

\begin{figure}
\centering
\includegraphics[width=9cm]{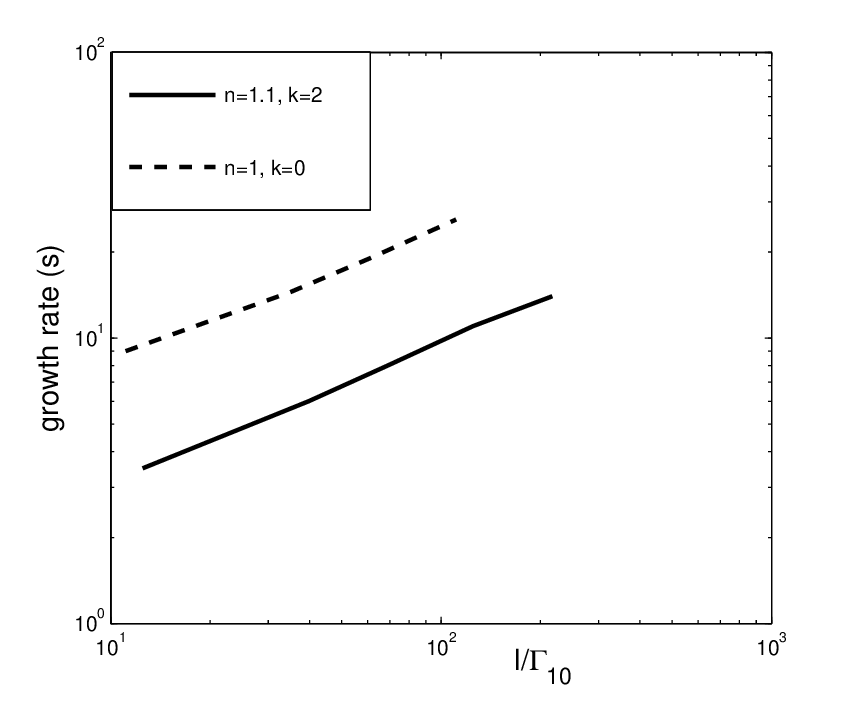}
\caption{\label{fig:g}Dimensionless growth rate versus $l/\Gamma_{10}$.  The solid line corresponds to $m=0.645$
and the dashed line to $m=2$.}
\end{figure}

For the simulations of non-spherical perturbations ($l\ne0$) we employed the analytic solution
given in (\ref{spher-mod-1})-(\ref{spher-mod-2}) as the initial condition.  We have also made some runs 
for comparison using the second spherical mode found above as the initial condition and verified that the long term evolution
of non-spherical growing modes is essentially independent of the details of the initial state.   The free parameters
of the model are: the indices $k$ and $n$ characterizing the density profiles of the unshocked ambient 
medium and ejecta (see figure \ref{fig:a}), and the adiabatic index $\hat{\gamma}$ of the shocked ejecta.  The 
exponent $m$ and the ratio $q$ are given by  (\ref{m}) and (\ref{q}), respectively, for a given 
choice of $n$, $k$ and $\hat{\gamma}$.  Since the reverse shock is non or at best mildly relativistic in the cases 
examined below we adopt $\hat{\gamma}=5/3$.  We have explored solutions for a range of values 
of $n$ and $k$ for which $m$ lies in the range $0.5-2$. 
The results displayed in figures \ref{fig:c}-\ref{fig:f} were computed using our canonical choice of parameters: $n=1.1$, $k=2$, for
which $m=0.645$, $q=1.06$.  The corresponding unperturbed solution is 
displayed in figure \ref{fig:b} (left panel).   For other values of $n$ and $k$ the solution exhibits 
the same qualitative behavior, albeit with a larger growth rate for larger values of $m$ .  
The dependence of the convective growth rate on $m$ is
examined in figure \ref{fig:g} and discussed below.

Quite generally we find that eigenmodes having angular scales larger than the causality scale, roughly
$l/\Gamma_1<1$, decay with time.  This is expected since for these modes information can only 
propagate over distances smaller than the characteristic wavelength, so that a causal section evolves like a spherical mode.
An example is shown in figure \ref{fig:c}, where solutions obtained for $l(l+1)/\Gamma_{10}^2=0.2$ (upper panels) and 
$l(l+1)/\Gamma_{10}^2=1.5$ (bottom panels) are exhibited. Here $\Gamma_{10}=\Gamma_1(\tau=0)$ denotes
the initial Lorentz factor of the forward shock.  The analytic solution for the spherical mode is plotted 
for a comparison in the right panels (dashed red line).  The evolution of the dimensionless distortions of the forward and 
reverse shock fronts, $\delta r_{1(2)}/r_{1(2)}=\delta_{1(2)}/\Gamma_{1(2)}^2$, and the contact surface,  
$\delta r_c/r_c=\delta_c/\Gamma_{c}^2$, (see equation (\ref{del_rs})) are displayed in the left panels. 
Relative pressure perturbations are shown in the right panels.  The decay of perturbations with 
an angular scale sufficiently larger than the initial causality scale is evident from this example.
The deviation from the analytic solution is small, as seen in the upper right panel.  Evolution of
longer wavelength modes, $l(l+1)/\Gamma_{10}^2<0.2$, is practically identical to that of the spherical mode.   
The turnover at $\tau\sim 1.5$ exhibited by the solution displayed in the bottom panels is due the fact 
that $l(l+1)/\Gamma_{1}^2$ increases with time as $\exp(m\tau)$, so that the wave enters the 'horizon' in 
this case early enough to affect the evolution of the perturbations at the contact surface.  This is roughly the 
boarder case separating stable and unstable modes.

Modes of order $l>\Gamma_{10}$ are found to be unstable.  This is demonstrated in figure \ref{fig:d}, where the
evolution of an eigenmode of order $l=10^2 \Gamma_{10}$ is delineated.  Oscillations resulting 
from sound waves crossing, followed by a rapid growth of perturbations in the shocked ejecta and near the contact 
in the shocked ambient medium are clearly seen.  The frequency of 
oscillations is found to be proportional to the spherical harmonic degree $l$, as anticipated.  Moreover, the frequency 
of oscillations in region 2 is smaller than that in region 1, owing to the difference in specific enthalpies of the 
fluids on each side of the contact surface.  The growth of perturbations at the contact 
surface commences very early on, as seen in figure \ref{fig:d}.  The instability
then propagates from the contact to the forward and reverse shocks.  The reverse shock responds
rather quickly, mainly because it is located much closer to the contact than the forward shock.
The propagation of the instability in the shocked ambient medium (region 1) is rather slow, as 
illustrated in figure \ref{fig:e}.  The reason is that energy pumped from the contact into this region is now distributed over a much 
larger volume.  However, in reality the instability near the contact will quickly reach the nonlinear regime 
and saturate, at which point the linear analysis breaks down.  What then anticipated
is formation of R-T fingers that expand to the forward shock, as seen in the nonrelativistic
case \citep{CBE92,JN96}.  To study the response of the forward shock requires full, high resolution 
MHD simulations.

The growth of the perturbations that starts following the decay of the transient initial state is
exponential in the time $\tau$, or algebraic in the physical time $t$, viz. $\delta Q\propto {\rm e}^{s\tau}\propto (t/t_0)^s$
where $Q$ represents any of the fluid quantities.  This is shown in  
figure \ref{fig:f} for the relativistic distortion $\delta_c$ defined in  (\ref{del_rs}).  Growth of variables located at larger distances
from the contact commences at later times, but follow with the roughly same growth rate.  It is also evident from
figure \ref{fig:f} that the growth rate $s$ increase with increasing mode degree $l$.  From our numerical simulations we 
find the scaling $s\propto \sqrt{l/\Gamma_{10}}$, with the proportionality constant depending predominantly
on $m$.  Examples are exhibited in figure \ref{fig:g} for two different cases; the first one corresponds to 
$m=0.645$ (solid line) and the second one to $m=2$ (dashed line).
A fit to the lines in fig \ref{fig:g} gives $s\simeq\sqrt{l/\Gamma_{10}}$ in the 
former case and $s\simeq2.45\sqrt{l/\Gamma_{10}}$ in the latter.
This scaling is expected for a R-T instability which is driven by the ``effective'' gravitational
force felt by the decelerating contact interface.  To see this,   
let $a_c^\mu$ denotes the 4-acceleration of the contact interface and $u_c^\mu$ its 4-velocity.  In flat spacetime the acceleration
is orthogonal to the velocity, $a^\mu u_\mu=0$.  In the rest frame of the contact $a^\mu=(0,a_r^\prime,0,0)$ from which 
we obtain $a^2=(a_r^{\prime})^2$.  From these relations and the normalization of the 4-velocity, $u^2=1$, we get 
$a_r^\prime={\rm d}u^r/{\rm d}t\simeq {\rm d}\Gamma_c/{\rm d}t=-m\Gamma_c/2t$ for $\Gamma^2_c\propto (t/t_0)^{-m}$.   The ``effective'' gravitational 
force felt by the decelerating contact is just $g=-a_r^\prime$.  Now, the wavevector component parallel
to the contact surface, $k_{||}$, is given roughly by $k_{||}=r_c/l$ for an eigenmode of order $l$.  We thus have 
$\sqrt{gk_{||}}t^\prime\simeq\sqrt{ml/2\Gamma_c}\propto s$, noting that $t^\prime=t/\Gamma_c$ and that 
the ratio $\Gamma_c/\Gamma_1$ is independent of time.  Consequently, the growth rate, as measured in the rest frame of the 
contact, is $s/t^\prime\propto \sqrt{gk_{||}}$, which is just the R-T growth rate in the case $\rho_{2c}\gg \rho_{1c}$.
Note that the dependence of $s$ on $m$, as derived from the simulations, reflects not only the effect of deceleration but also 
the dependence of the unperturbed solution on $m$, in particular the density ratio at the contact, as can be seen from figure \ref{fig:b}.

\section{\label{sec:imp}Implications for GRBs}
In the standard GRB model the afterglow emission observed following the explosion is commonly attributed to synchrotron 
cooling of relativistic electrons behind the forward shock \citep{MzR97} (but c.f., \citep{UB07,GDM07}).  
The observations seem to indicate the presence of strong magnetic fields over a significant portion of the shocked circumburst layer. 
The magnetic energy density estimated from the data, roughly a fraction $\epsilon_B\sim10^{-3}-10^{-1}$ of the internal
energy density \citep{PK02}, is several orders of magnitudes larger than that expected from compression of 
the preshock magnetic field. 
This implies magnetic field generation or amplification by some mechanism.  Whether kinetic processes can generate such magnetic 
fields is yet an open issue.  Plasma instabilities that develop in the collisionless shock transition generate strong
magnetic fields on kinetic scales.  However, recent shock simulations \citep{Sp07} indicate that the fields thereby produced 
decay rapidly over a few skin depths, before reaching the MHD scale.  
Magnetic field generation by streaming ultra-relativistic protons (or cosmic rays) in the immediate shock upstream has 
also been considered.  In this scenario the magnetized cosmic ray precursor is envisioned to form as an inherent part 
of the shock transition.  The key question here is whether the shock can indeed inject enough energy in the form of ultra-relativistic 
particles to sustain the required large scale magnetic field.   To study this process using self-consistent shock simulations requires
computing time beyond present capabilities. 

An alternative to plasma instabilities is magnetic field amplification by MHD turbulence .  It has been proposed \citep{SG07} that 
macroscopic turbulence might be produced via the interaction of the forward shock with a clumpy circumburst medium.
In this scenario the response of the shock to the preshock density inhomogeneities leads to generation of vorticity and the consequent
amplification of magnetic fields \citep{ZM09}.   Amplification of magnetic energy to the level inferred from observations requires 
large (order unity) density contrasts and filling factors of the clumps.   Whether such conditions exist in the surrounding environment 
is yet an open issue.   Here we propose that the convective instability found above may be an inherent source of turbulence in
the shocked circumburst layer, at least at early times.  The instability may also lead to nonlinear distortions of the
shock front itself without the need for an external driver.   If the ejecta is magnetized at a level smaller than that required 
to suppress the instability but still 
much larger than that of the unshocked ambient medium, then mixing of the magnetized ejecta with the shocked ambient gas via growth 
of R-T fingers alone can lead to strong magnetization of shocked cirumburst layer at sufficiently early times.
How the system evolves at later times, after the reverse shock crosses the ejecta, is unclear at present.    
The stability analysis of the BMK solution performed by Gruzinov \citeyearpar{G00} suggests that 
it may be a very slow attractor.  Linear perturbations
of the forward shock in the BMK phase decay very slowly.  Whether this behavior persists also 
in the nonlinear regime is yet to be determined.  If it does then it could well be that the growth of R-T fingers and, perhaps, 
the nonlinear oscillations of the forward shock itself which are induced by the convective instability may be a source of vorticity
during a long portion of the evolution of the blast wave \citep{MN07}.  
The simulations performed by Zhang et al. \citeyearpar{ZM09} suggest a saturation level of $\epsilon_B\sim5\times10^{-3}$
for the turbulence induced magnetic energy density, weakly dependent on the initial magnetic field strength. 

The convective instability may also affect the emission from the shocked ejecta. As shown above, 
the effect of the instability on the reverse shock is prompt and dramatic.
The nonlinear distortions of the reverse shock may strongly alter
particle acceleration and the emission processes during reverse shock crossing.  It is tempting to speculate that 
the lack of observed optical flashes, that are anticipated in the ``standard'' model, and the 
behavior of the early afterglow phase may be attributed to the instability discussed here, although we do not
offer at present any specific explanation.  At any rate, the salient lesson is that a careful 
analysis that takes account of this process is required to better understand the observational characteristics
of the emission during the early post-prompt phase.

\section{\label{sec:conc}Summary}
We have performed a global linear stability analysis of a self-similar solution describing the interaction of a 
relativistic shell with an ambient medium.   Our analysis indicates a strong convective instability at early 
stages of the evolution of the dense ejecta as it sweeps a lighter ambient gas.  Our main findings are:

1. Eigenmodes having angular scales smaller than the causality scale, roughly $l/\Gamma_1>1$, where $\Gamma_1$ 
is the Lorentz factor of the blast wave, are unstable and exhibit a rapid growth.   
Lower order modes for which $l/\Gamma_1<1$ are stable.  

2. Growth of perturbations starts promptly near the contact discontinuity.  The instability then propagates
towards the forward and reverse shocks.  The reverse shock responds quickly to the growth of distortions at the 
contact.   Propagation of the signal to the forward shock is much slower.  The instability near the contact becomes 
nonlinear well before the signal arrives at 
the forward shock, so full MHD simulations are needed to resolve the effect of the instability on the forward shock.

3. The growth is algebraic in time, that is, $\delta Q\propto (t/t_0)^s$ for any fluid quantity $Q$.  
The dimensionless growth rate scales as $s\propto\sqrt{l/\Gamma_1}$,
with a proportionality constant that increases with increasing $m$.  This implies development of a very small
scale structure with a significant amplitude, up to the dissipation scale, at least at sufficiently early times.  
The effect of such corrugations on the collisionless shock transition and related processes, particularly 
particle acceleration, needs to be explored.

Unfortunately, the linear analysis outlined above is restricted to a limited set of conditions under which the 
unperturbed self-similar solution of Nakamura \& Shigeyama \citeyearpar{NS06} is applicable.  
It is naively expected that the instability will be strongly suppressed in cases where the ejecta is highly 
magnetized and/or if the reverse shock is highly relativistic.   
Full 3D MHD simulations should be exploited to study this process in other situations, and to follow the  
evolution of the convective instability in the nonlinear regime.  As illustrated above, high resolution 
simulations that can resolve angular scales $\Delta \theta<<1/\Gamma$ are required, posing a great numerical challenge.
We believe that our findings strongly motivate such efforts.

\vspace{12pt}

I thank A. Ditkowski for a technical help in the development of the code,
and M. Aloy, D. Kushnir, M. Lyutikov, A. MacFadyen, E. Nakar and E. Waxman for enlightening discussions.
This work was supported by an ISF grant for the Israeli Center for High Energy Astrophysics,
and by the NORDITA program on Physics of relativistic flows.

\newpage

\appendices
\section{\vspace{12pt}\\Derivation of the unperturbed flow equations}
\subsection{\label{sec:app_unp1}Region 1}
The derivation of the flow equations in region 1 (the shocked ambient medium) 
is straightforward and follows that in \cite{BMK76}.
Transforming from the coordinates $(r,t)$ to $(\chi,\tau)$, with $\tau=\ln t$ and $\chi$ 
given by (\ref{chi}), and using the relations
\bme
\se
\label{dt2}
\begin{eqnarray}
t\partial_t&=&\partial_\tau+[(m+1)(2\Gamma_1^2-\chi)+1]\partial_\chi,\\
t\partial_r&=&-[1+2(m+1)\Gamma_1^2]\partial_\chi,\\
t{\rm d}^0_t&=&\partial_\tau+(m+1)(2/g-\chi)\partial_\chi,
\end{eqnarray}
\eme
one obtains, upon substitution of equations (\ref{g}a-c) into the flow equations (\ref{momentum}a,b),
\bme
\se
\label{out}
\begin{eqnarray}
\frac{1}{g}\frac{{\rm d}\ln g}{{\rm d}\chi}&=& \frac{(7m+3k-4)-(m+2)g\chi}{(m+1)(4-8g\chi+g^2\chi^2)},\\
\frac{1}{g}\frac{{\rm d}\ln f}{{\rm d}\chi}&=&\frac{8(m-1)+4k-(m+k-4)g\chi}{(m+1)(4-8g\chi+g^2\chi^2)},\\
\frac{1}{g}\frac{{\rm d}\ln h}{{\rm d}\chi}&=&\frac{2(9m+5k-8)-2(5m+4k-6)g\chi+(m+k-2)g^2\chi^2}
{(m+1)(4-8g\chi+g^2\chi^2)(2-g\chi)}.
\end{eqnarray}
\eme

\subsection{\label{sec:app_unp2}Region 2}
In region 2 we use the coordinates $(\tau,\sigma)$. Then,
\bme
\se
\label{dt2-inn}
\begin{eqnarray}
t\partial_t&=&\partial_\tau+[(m+1)(2\Gamma_2^2-\sigma)+1]\partial_\sigma,\\
t\partial_r&=&-[1+2(m+1)\Gamma_2^2]\partial_\sigma,\\
t{\rm d}^0_t&=&\partial_\tau+(m+1)(1/qG-\sigma)\partial_\sigma.
\end{eqnarray}
\eme
Substituting  (\ref{G2}a-c) into equations (\ref{momentum}a,b) yields
\bme
\se
\label{in}
\begin{eqnarray}
2(1+qG\sigma)\frac{{\rm d}\ln F}{{\rm d}\sigma}-(1-qG\sigma)\kappa\frac{{\rm d}\ln G}{{\rm d}\sigma}&=&
\frac{(mn-\kappa m-6)}{(m+1)}qG,\\
2(1-qG\sigma)\frac{{\rm d}\ln F}{{\rm d}\sigma}-\hat{\gamma}(1+qG\sigma)\frac{{\rm d}\ln G}{{\rm d}\sigma}&=&
\frac{[\hat{\gamma}(m+2)-mn-6(\hat{\gamma}-1)]}{(m+1)}qG,\\
2(1-qG\sigma)\frac{{\rm d}\ln H}{{\rm d}\sigma}-2\frac{{\rm d}\ln G}{{\rm d}\sigma}&=&-\frac{(mn-m-2)}{(m+1)}qG,
\end{eqnarray}
\eme
where $\kappa(\sigma)=W_2/p_2\gamma_2^2$, and $\hat{\gamma}$ denotes the adiabatic index.

\newpage

\section{\vspace{12pt}\\Derivation of the equations for the dimensionless perturbations}
\subsection{\label{sec:app_r1}Region 1}
To derive the equations for the dimensionless perturbations $\xi_\alpha(\tau,\chi)$ we first write 
equations (\ref{prt}a-d) in the new coordinates $(\chi,\tau)$, using the 
relations (\ref{dt2}a-c) and (\ref{xirho}a-d).  Recalling that $W_1=4p_1\gamma_1^2$
and noting that $r\nabla\delta{\bf v}_{1T}=-l(l+1)\xi_TY_{l\tilde{m}}$, we obtain from the continuity equation
(\ref{prt}a)
\bme
\se
\label{xi1}
\begin{equation}
\partial_\tau\xi_\rho+(m+1)(2/g-\chi)\partial_\chi\xi_\rho-\frac{2(m+1)}{g}\partial_\chi\xi_R
+\frac{2(m+1)}{g}(\partial_\chi\ln g-\partial_\chi\ln h)\xi_R=l(l+1)\xi_T.
\end{equation}
Equation (\ref{prt}b) gives
\begin{eqnarray}
\partial_\tau(\xi_P+\frac{2}{3}\xi_R)+\frac{(m+1)}{g}[(2-g\chi)\partial_\chi\xi_P-
\frac{2}{3}(2+g\chi)\partial_\chi\xi_R]\label{xi2}\,&&\nonumber\\
+\frac{2(m+1)}{g}[\frac{2}{3}\partial_\chi\ln g
-\partial_\chi\ln f]\xi_R&=&\frac{4}{3}l(l+1)\xi_T,
\end{eqnarray}
the transverse component of the momentum equation (\ref{prt}c)) gives
\begin{equation}
\partial_\tau\xi_T+(m+1)(2/g-\chi)\partial_\chi\xi_T+[(m+1)(2/g-\chi)
\partial_\chi\ln(fg/h)-m]\xi_T=-\frac{\xi_P}{2g\Gamma^2},
\end{equation}
and the energy equation (\ref{prt}d) gives
\begin{equation}
\partial_\tau(\xi_P+2\xi_R)+2(m+1)(2/g-\chi)\partial_\chi\xi_R-(m+1)(2/g+\chi)\partial_\chi\xi_P
+c_{oR}\xi_R+c_{oP}\xi_P=0,
\end{equation}
\eme
where
\bme
\se
\begin{eqnarray}
c_{oR}&=&-(3m+k)-(m+1)\chi\partial_\chi[2(\ln g)+(\ln f)],\\
c_{oP}&=&-(3m+k)+2(m+1)(2/g-\chi)\partial_\chi(\ln g)-(m+1)(2/g+\chi)\partial_\chi(\ln f).
\end{eqnarray}
\eme
Note that (\ref{xi1}c) holds only for $l\ne0$.  In this case it is readily seen from (\ref{xi1}b,c) 
that $\xi_P$ and $\xi_T$ must have different time evolution, owing to the extra factor $\Gamma^{-2}$ on the right-hand side of
(\ref{xi1}c), implying breakdown of self-similarity.  In the spherical case ($l=0$), $\nabla_T Y_{l\tilde{m}}=0$, 
and equation (\ref{prt}c) is identically zero.  The right-hand sides of (\ref{xi1}a,b) then vanishes allowing 
separation of variables.

For an impulsive BMK76 solution with $k=0$, $m=3$, $g=\chi^{-1}$, $h=\xi^{-7/4}$ and 
$f=\chi^{-17/12}$ the above equations reduce to
\bme
\se
\label{bmk}
\begin{eqnarray}
\partial_\tau\xi_\rho+4\chi\partial_\chi\xi_\rho-8\chi\partial_\chi\xi_R+6\xi_R&=&l(l+1)\xi_T,\\
\partial_\tau[\xi_P+(2/3)\xi_R]+4\chi\partial_\chi\xi_P-8\chi\partial_\chi\xi_R+6\xi_R
&=&\frac{4}{3}l(l+1)\xi_T,\\
\partial_\tau\xi_T+4\chi\partial_\chi\xi_T-\frac{17}{3}\xi_T+\frac{\xi_P}{2g\Gamma^2}&=&0,\\
\partial_\tau(\xi_P+2\xi_R)+8\chi\partial_\chi\xi_R-12\chi\partial_\chi\xi_P+\frac{14}{3}\xi_R&=&0,
\end{eqnarray}
\eme
as derived originally by Gruzinov \citeyearpar{G00} (except for the 3rd term on the right-hand side of (\ref{bmk}c), see
footnote at the end of section \ref{sec:pert_scheme}).
After some manipulation of equations (\ref{xi1}a-d) we arrive at
(\ref{xiset}), with the different coefficients given by
\bme
\se
\label{ARR}
\begin{eqnarray}
A_{RR}&=&A_{PP}=(m+1)(\chi-4/g),\\
A_{PR}&=&\frac{4}{3}A_{RP}=2A_{\rho R}=\frac{4(m+1)}{g},\\
A_{\rho \rho}&=&A_{TT}=-(m+1)(2/g-\chi),
\end{eqnarray}
\eme
and
\bme
\se
\label{BRR}
\begin{eqnarray}
B_{RR}&=&\frac{3(3m+k)}{4}+\frac{(m+1)}{2}\left(\frac{2}{g}+3\chi\right)\partial_\chi\ln g
-\frac{3(m+1)}{4}\left(\frac{2}{g}-\chi\right)\partial_\chi\ln f,\\
B_{PR}&=&-\frac{(3m+k)}{2}-(m+1)\left(\frac{2}{g}+\chi\right)\partial_\chi\ln g
+\frac{(m+1)}{2}\left(\frac{6}{g}-\chi\right)\partial_\chi\ln f,\\
B_{PP}&=&-\frac{3}{2}B_{RP}=-\frac{(3m+k)}{2}+(m+1)\left(\frac{2}{g}-\chi\right)\partial_\chi\ln g
-\frac{(m+1)}{2}\left(\frac{2}{g}+\chi\right)\partial_\chi\ln f,\\
B_{PT}&=&-2B_{RT}=2B_{\rho T}=2l(l+1),\\
B_{\rho R}&=&-\frac{2(m+1)}{g}\partial_\chi\ln(g/h),\\
B_{TP}&=&-1/(2g\Gamma^2),\\
B_{TT}&=&m-(m+1)\left(\frac{2}{g}-\chi\right)\partial_\chi\ln(fg/h).
\end{eqnarray}
\eme
All other coefficients vanish.
\subsection{\label{sec:app_r2}Region 2}
The derivation of the perturbation equations in region 2 is similar to that in region 1, but slightly more involved.
Here we express (\ref{prt}a-d) in the coordinates $(\sigma,\tau)$, using the 
relations (\ref{dt2-inn}a-c) and (\ref{etarho}a-d). 
To shorten the notation we define  $a=\hat{\gamma}/(\hat{\gamma}-1)$ and $\Delta_\pm=(m+1)(1/qG\pm\sigma)$.
From the continuity equation (\ref{prt}a) we then obtain
\bme
\se
\label{eta}
\begin{equation}
\partial_\tau\eta_\rho+\Delta_-\partial_\sigma\eta_\rho-\frac{2(m+1)}{G}\partial_\sigma\eta_R
+\frac{2(m+1)}{G}(\partial_\sigma\ln G-\partial_\sigma\ln H)\eta_R=l(l+1)\eta_T,
\end{equation}
noting that $r\nabla\delta{\bf v}_{2T}=-l(l+1)\eta_TY_{l\tilde{m}}$.
Likewise, (\ref{prt}b) yields
\begin{equation}
\partial_\tau\eta_T+\Delta_-\partial_\sigma\eta_T+\left[\frac{a}{\kappa}\Delta_-\partial_\sigma\ln(F/H)
+\frac{\kappa+a}{2\kappa}\Delta_-\partial_\sigma\ln G-m/2\right]\eta_T+\frac{\eta_P}{\kappa qG\Gamma_2^2}=0,
\end{equation}
where the relation $W_{20}/p_{20}=\kappa qG\Gamma_2^2$ has been used, 
the transverse component of the momentum equation (\ref{prt}c) gives
\begin{equation}
\partial_\tau(\eta_P+\hat{\gamma}q\eta_R)+\Delta_-\partial_\sigma\eta_P-
\hat{\gamma}q\Delta_+\partial_\sigma\eta_R
+\frac{(m+1)}{G}[\hat{\gamma}\partial_\sigma\ln G
-2\partial_\sigma\ln F]\eta_R=\hat{\gamma}l(l+1)\eta_T,
\end{equation}
and (\ref{prt}d) gives
\begin{equation}
\kappa q\partial_\tau\eta_R+\partial_\tau\eta_P+\kappa q\Delta_-\partial_\sigma\eta_R -\Delta_+
\partial_\sigma\eta_P+c_{iR}\eta_R+c_{iP}\eta_P+c_{i\rho}\eta_\rho=0,
\end{equation}
\eme
with
\bme
\se
\begin{eqnarray}
c_{i\rho}&=&-c_{ip}=\frac{\kappa-a}{2}(-m+\Delta_{-}\partial_\sigma\ln G),\\
c_{iR}&=&\frac{2(m+1)}{G}\partial_\sigma\ln F-\kappa\frac{ (m+1)}{G}\partial_\sigma\ln G-qc_{i\rho}.
\end{eqnarray}
\eme

After some algebra equations (\ref{eta}a-d) can be recast into the form of
(\ref{etaset}), with the coefficients $C_{\alpha\beta}$, $D_{\alpha\beta}$ given explicitly by:
\bme
\se
\label{C1}
\begin{eqnarray}
C_{RR}&=&C_{PP}=(m+1)\left[\sigma-\frac{\kappa+\hat{\gamma}}{qG(\kappa-\hat{\gamma})}\right],\\
C_{RP}&=&\frac{1}{q^2\hat{\gamma}\kappa}C_{PR}=\frac{2(m+1)}{q^2G(\kappa-\hat{\gamma})},\\
C_{\rho R}&=&\frac{2(m+1)}{G},\\
C_{\rho\rho}&=&C_{TT}=(m+1)(\sigma-1/qG),
\end{eqnarray}
\eme
and
\bme
\se
\label{DTT}
\begin{eqnarray}
D_{RR}&=&\frac{(m+1)}{qG(\kappa-\hat{\gamma})}\left[-4\partial_\sigma\ln F
+(\kappa+\hat{\gamma})\partial_\sigma\ln G\right]
+\frac{\kappa-a}{2(\kappa-\hat{\gamma})}[\Delta_-\partial_\sigma\ln G-m],\\
D_{PR}&=&\frac{2(m+1)}{G(\kappa-\hat{\gamma})}\left[(\hat{\gamma}+\kappa)\partial_\sigma\ln F
-\kappa\hat{\gamma}\partial_\sigma\ln G\right]
-\frac{q\hat{\gamma}(\kappa-a)}{2(\kappa-\hat{\gamma})}[\Delta_-\partial_\sigma\ln G-m],\\
D_{PP}&=&-q\hat{\gamma}D_{RP}=-\frac{\hat{\gamma}(\kappa-a)}{2(\kappa-\hat{\gamma})}
[\Delta_-\partial_\sigma\ln G-m],\\
D_{PT}&=&-q\kappa D_{RT}=\frac{\hat{\gamma}\kappa l(l+1)}{(\kappa-\hat{\gamma})q},\\
D_{\rho R}&=&\frac{2(m+1)}{G}\partial_\sigma\ln(G/H),\\
D_{\rho T}&=&l(l+1),\\
D_{TP}&=&-\frac{1}{\kappa q\Gamma_2^2 G},\\
D_{TT}&=&-\frac{a}{\kappa}\Delta_{-}\partial_\sigma\ln(F/H)-\frac{\kappa+a}{2\kappa}
\Delta_{-}\partial_\sigma\ln G+m/2.
\end{eqnarray}
\eme

\section{\label{sec:app_BC}\vspace{12pt}\\Boundary conditions at the reverse shock}
The normal to the reverse shock is written $n_{2\mu}=n_{2\mu}^0+\delta n_{2\mu}$, where $n_{2\mu}^0$ 
and $\delta n_{2\mu}$ are given by equations (\ref{pert_normal}a,b), respectively, with the subscript 
1 replaced by 2.  Applying  (\ref{cont-s}a) to the reverse shock we have
\begin{equation}
\rho_{20}^\prime v_{20}^\mu\delta n_{2\mu}+(\Delta_2 \rho_2^\prime v^\mu_{20}+\rho_{20}^\prime\Delta_2 v^\mu_2)n^0_\mu
=\rho_{e}^\prime v_{e}^\mu\delta n_{2\mu}+(\Delta_2 \rho_e^\prime v^\mu_{e}+\rho_{e}^\prime\Delta_2 v^\mu_e)n^0_\mu,
\label{bc-2-rho}
\end{equation}
with the density of the unshocked ejecta $\rho_e$ is given by  (\ref{den-ejecta}), $v_e^\mu=(1,v_e,0,0)$, etc.
The velocity of the unshocked ejecta near the shock surface evolves with time 
as ${\rm d}v_e/{\rm d}t=\partial_t v_e+(V_2+\delta V_2)\partial_r v_e$.
Integrating over time from $t=t_0$ we find $\Delta_2 v_e=(\delta r_2/t)(1-\delta r_{20}/\delta r_2)$ to the order 
to which we are working, where $\delta r_{20}$ is the initial displacement of the reverse shock surface.  From   (\ref{del_rs})
we then obtain $\Delta_2 v_e^\mu n_{2\mu}^0=(1-\delta r_{20}/\delta r_2)(\delta_2/\Gamma_2)Y_{l\tilde{m}}$.
The corresponding convective change of the density $\rho_e^\prime$ is $ \Delta_2\ln\rho_e^\prime
=(1-n)(m+1)\Gamma_2^2\Delta_2 v_e $.
Using  (\ref{pert_normal}b) with the subscript 1 replaced by 2 we also have
$2(m+1)v_{e}^\mu\delta n_{2\mu}=-(m+2)\Gamma_2\delta V_2 Y_{l\tilde{m}}$.  For the perturbations of the
shocked ejecta we obtain, using the unperturbed flow parameters (\ref{G2}a-c),
$\Delta_2 \ln\rho_2^\prime=[\eta_\rho-2(m+1)\delta_2\partial_\sigma(\ln H)_2]Y_{l\tilde{m}}$, and
$q\Gamma^2_2\Delta_2 v_2=[q\eta_R-(m+1)\delta_2\partial_\sigma(\ln G)_2]Y_{l\tilde{m}}$.
Substituting the above results into  (\ref{bc-2-rho}) yields
\begin{eqnarray}
\partial_\tau\delta_2+A_{sR}\eta_R+A_{s\rho}\eta_\rho+A_{s\delta}\delta_2=0,\label{bc-2-app1}
\end{eqnarray}
where
\bme
\se
\label{Asdel}
\begin{eqnarray}
A_{sR}&=&\frac{2q}{q-1}A_{s\rho}=\frac{qm}{q-m-1},\\
A_{s\delta}&=&m+1-\frac{m(m+1)}{(q-m-1)}\{\partial_\sigma(\ln G)_2+(q-1)\partial_\sigma(\ln H)_2\}\nonumber\\
&&\,-\frac{(m+1)(m+2-mn)(q-1)}{2(q-m-1)}(1-\delta r_{20}/\delta r_{2}).
\end{eqnarray}
\eme
The transverse component of  (\ref{cont-s}b) reduces to
\begin{equation}
\eta_T=\frac{2}{\kappa(q-1)}\frac{\delta_2}{\Gamma_2^2},\label{bc-2-app2}
\end{equation}
and the zeroth component reads
\begin{eqnarray}
\label{inner-zero}
W_{20}v_{20}^\mu\delta n_{2\mu}+(\Delta_2 W_{2}v_{20}^\mu+W_{20}\Delta_2 v_{2}^\mu)n_{2\mu}&-&
(p_{20}\delta n_{20}+\Delta_2 p_{2}n_{20})=\nonumber\\[0.2em] 
&&W_{e}v_{e}^\mu\delta n_{2\mu}+(\Delta_2 W_{e}v_{e}^\mu+W_{e}\Delta_2 v_{e}^\mu)n_{2\mu},
\end{eqnarray}
with $W_e=\rho_e\gamma_e^2$, $\Delta_2 W_e=\rho_e\gamma_e^2(2-n)(m+1)\delta_2 (1-\delta r_{20}/\delta r_2)Y_{l\tilde{m}}$,
$\Delta_2 W_2=p_{20}(\kappa+a)q^2(\Gamma_4^2\Delta v_2)+q(\kappa-a)\Delta \ln\rho_2^\prime+qa\Delta_2 \ln p_2$,
and $\Delta_2 \ln p_2=[\eta_P-2(m+1)\delta_2\partial_\sigma(\ln F)_2]Y_{l\tilde{m}}$.
Rearranging terms we finally arrive at
\begin{equation}
B_{st}\partial_\tau\delta_2+B_{s\delta}\delta_2+B_{sR}\eta_R+B_{sP}\eta_P+B_{s\rho}\eta_{\rho}=0,\label{bc-2-app3}
\end{equation}
where
\bme
\se
\label{Bsdel}
\begin{eqnarray}
B_{st}&=&1-\frac{q+1}{2}\kappa+\frac{m+2}{2}\frac{\rho_e}{P_{20}},\\
B_{sR}&=&\frac{q(q+1)}{2}(\kappa-a)+aq^2,\\
B_{sP}&=&\frac{a(q-1)+2}{2},\\
B_{s\rho}&=&\frac{q-1}{2}(\kappa-a),\\
B_{s\delta}&=&(m+1)\left[B_{st}-\frac{B_{sR}}{q}\partial_\sigma(\ln G)
-2B_{sP}\partial_\sigma(\ln F)-2B_{s\rho}\partial_\sigma(\ln H)\right] \nonumber\\
&&\,+\frac{\rho_e}{P_{20}}(m+1)\left(mn/2-m-1\right)(1-\delta r_{20}/\delta r_{2}).
\end{eqnarray}
\eme

Subtracting the radial component of (\ref{cont-s}b) from (\ref{inner-zero}) we obtain
\begin{eqnarray}
[W_{20}v_{20}^\mu\delta n_{2\mu}+(\Delta_2 W_{2}v_{20}^\mu+W_{20}\Delta_2 v_{2}^\mu)n^0_{2\mu}](v_{20}-v_e)
+W_{20}(\Delta_2 v_2) v_{20}^\mu n^0_{2\mu}\,&&\nonumber\\
+p_{20}(V_2-v_e)\Gamma_2^3\delta V_2+\Delta_2 p_{2}(1-V_2v_e)\Gamma_2&=&W_e(\Delta_2 v_e) v_e^\mu n^0_{2\mu},
\end{eqnarray}
whereby the final condition, 
\begin{equation}
C_{st}\partial_\tau\delta_2+C_{s\delta}\delta_2+C_{sR}\eta_R+C_{sP}\eta_P+C_{s\rho}\eta_{\rho}=0,\label{bc-2-app4}
\end{equation}
is derived, with
\bme
\se
\label{Cs}
\begin{eqnarray}
C_{st}&=&\frac{(m+1-q)(q+1)\kappa}{2q}-m,\\
C_{sR}&=&\frac{(q-m-1)}{2}[(q+1)\kappa+(q-1)a]+(q-1)(m+1)\kappa,\\
C_{sP}&=&\frac{(q-m-1)(q-1)}{2q}a+(m+2),\\
C_{s\rho}&=&\frac{(q-m-1)(q+1)}{2q}(\kappa-a),\\
C_{s\delta}&=&(m+1)\left[C_{st}-\frac{C_{sR}}{q}\partial_\sigma(\ln G)-2C_{sP}\partial_\sigma(\ln F)-2C_{s\rho}
\partial_\sigma(\ln H)\right].
\end{eqnarray}
\eme
With the exception of the spherical mode associated with a linear time translation of the solution outlined 
in section \ref{sec:unpert}, 
for which $\delta V_2=0$ and $\delta r_2(t)=\delta r_{20}$ (see section \ref{sec:pert_sph}), we shall assume $\delta r_{20}=0$.
We can then eliminate $\delta_2$ from  (\ref{bc-2-app1}),(\ref{bc-2-app2}),(\ref{bc-2-app3}) and 
(\ref{bc-2-app4}) to obtain equations (\ref{bc-eta}a-d), where we define
\bme
\se
\label{cof-2}
\begin{align}
d_{s\alpha}&=\frac{C_{s\alpha}B_{st}-B_{s\alpha}C_{st}}{C_{s\delta}B_{st}-B_{s\delta}C_{st}};&
\alpha&=(R,P,\rho),\\
e_{s\alpha}&=-\frac{C_{s\alpha}B_{s\delta}-B_{s\alpha}C_{s\delta}}{C_{s\delta}B_{st}-B_{s\delta}C_{st}};&
\alpha&=(R,P,\rho),\\
f_{s\lambda}&=\frac{B_{s\lambda}-A_{s\lambda}B_{st}}{B_{s\rho}-A_{s\rho}B_{st}};&
\lambda&=(R,P,\delta).
\end{align}
\eme

\section{\label{sec:app_RI}\vspace{12pt}\\Equations for the Riemann invariants}
Define the vector $|\xi\rangle=(\xi_R,\xi_P,\xi_\rho,\xi_T)$. Then (\ref{xiset}) can be expressed in matrix notation as 
\begin{equation}
\partial_\tau |\xi\rangle=A\partial_\sigma |\xi\rangle+B|\xi\rangle,
\end{equation}
with $A(\chi)$ and $B(\chi)$ denoting the matrices $(A_{\alpha\beta})$ and $(B_{\alpha\beta})$, respectively.
Let $|\psi_q\rangle$; $q=(+,-,3,4)$ be the eigenvectors of $A$ and $\lambda_q$ the corresponding eigenvalues.
We find 
\bme
\se
\begin{eqnarray}
\lambda_\pm&=&A_{RR}\pm\frac{\sqrt{3}}{2}A_{PR},\\
\lambda_3&=&\lambda_4=-(m+1)(2/g-\chi),\\
\end{eqnarray}
\eme 
where $A_{RR}$ and $A_{PR}$ are given by (\ref{ARR}a,b) respectively.
The Riemann invariants in region 1, given explicitly in (\ref{Rm-out}), are the coefficients of the expansion 
of $|\xi\rangle$ in the basis vectors $|\psi_q\rangle$, that is, $|\xi\rangle=\Sigma_q\xi_q|\psi_q\rangle$.  They obey
the equations
\begin{equation}
\partial_\tau\xi_q=\lambda_q\partial_\chi\xi_q+\Sigma_p{R_{qp}\xi_p},\label{Riemn1}
\end{equation}
where $R_{qp}=\langle\psi_q|B|\psi_p\rangle$.  We find that $\lambda_{-},\lambda_{3},\lambda_{4}$ are negative
everywhere in region 1, whereas $\lambda_{+}$ is positive, so that $\xi_-,\xi_3,\xi_4$ propagate
from the forward shock to the contact while $\xi_+$ propagates in the opposite direction.

The analysis in region 2 is similar but slightly more complicated, as the eigenvectors $|\zeta_p\rangle>
$ of the 
matrix $C$, the elements of which are given in  (\ref{C1}a-d), depend on the self-similarity 
parameter $\sigma$. The Riemann invariants defined in (\ref{Rm-in}) were computed
using the relation $|\eta\rangle=\Sigma_q\eta_q|\zeta_q\rangle$, where  $|\eta\rangle=(\eta_R,\eta_P,\eta_\rho,\eta_T)$.
For the eigenvalues of the matrix $C$ we have
\bme
\se
\begin{eqnarray}
\tilde{\lambda}_\pm&=&C_{RR}\pm \frac{C_{PR}}{q\sqrt{\kappa\hat{\gamma}}},\\
\tilde{\lambda}_3&=&\tilde{\lambda}_4=-(m+1)(1/qG-\sigma).
\end{eqnarray} 
\eme
The equation governing the evolution of the Riemann invariants in this region reads:
\begin{equation}
\partial_\tau\eta_q=\tilde{\lambda}_q\partial_\chi\eta_q+\Sigma_p{T_{qp}\eta_p},\label{Riemn2}
\end{equation}
where $T_{qp}=\langle\zeta_q|D|\zeta_p\rangle+\langle\zeta_q|C|\partial_\sigma\zeta_p\rangle$.  The eigenvalues 
$\tilde{\lambda}_{+},\tilde{\lambda}_{3},\tilde{\lambda}_{4}$ are positive everywhere 
while $\tilde{\lambda}_-$ is negative hence
$\eta_+,\eta_3,\eta_4$ propagate from the reverse shock to the contact while $\eta_-$ propagates 
from the contact to the reverse shock.

\end{document}